\documentclass{emulateapj}

\usepackage{verbatim}
\newcommand{\unit}[1]{\,{\rm #1}}
\newcommand{\pk}[1]{\ensuremath{k_{\rm pk}L/2\pi = #1}}
\newcommand{\edot}{\ensuremath{\dot{E}/\bar{\rho}L^2c_s^3}}

\begin{document}

\title{Dissipation and Heating in Supersonic Hydrodynamic and MHD Turbulence}
\author{M. Nicole Lemaster and James M. Stone}
\affil{Department of Astrophysical Sciences, Princeton University,
Princeton, NJ 08544}
\email{Lemaster@astro.princeton.edu}

\shorttitle{Dissipation and Heating in Turbulence}
\shortauthors{Lemaster \& Stone}

\begin{abstract}

We study energy dissipation and heating by supersonic MHD turbulence
in molecular clouds using Athena, a new higher-order Godunov code.  We
analyze the dependence of the saturation amplitude, energy dissipation
characteristics, power spectra, sonic scaling, and indicators of
intermittency in the turbulence on factors such as the magnetic field
strength, driving scale, energy injection rate, and numerical
resolution.  While convergence in the energies is reached at moderate
resolutions, we find that the power spectra require much higher
resolutions that are difficult to obtain.  In a $1024^3$ hydro run, we
find a power law relationship between the velocity dispersion and the
spatial scale on which it is measured, while for an MHD run at the
same resolution we find no such power law.  The time-variability and
temperature intermittency in the turbulence both show a dependence on
the driving scale, indicating that numerically driving turbulence by
an arbitrary mechanism may not allow a realistic representation of
these properties.  We also note similar features in the power spectrum
of the compressive component of velocity for supersonic MHD turbulence
as in the velocity spectrum of an initially-spherical MHD blast wave,
implying that the power law form does not rule out shocks, rather than
a turbulent cascade, playing a significant role in the regulation of
energy transfer between spatial scales.

\end{abstract}


\section{Introduction}

Observed non-thermal line widths in molecular clouds (MCs), where all
star formation in the Galaxy takes place, point to the presence of
supersonic turbulence in such regions (Falgarone \& Philips 1990).
The properties of the turbulent medium, such as Mach number and
magnetic field strength, may determine the products of the star
formation process.  As an important source of heating within
molecular clouds (Stone et al.~1998, hereafter S98),
turbulent energy dissipation may also play a role.  Much effort has
been directed toward numerically simulating turbulent media in order
to better understand the link between turbulence and star formation
(see Elmegreen \& Scalo 2004, MacLow \& Klessen 2004, and McKee \&
Ostriker 2007 and references therein).

As we have not yet identified the turbulent driving mechanism, there
remain many unanswered questions about the evolution of molecular
clouds.  Is the turbulence periodically re-energized, or does it
simply decay away?  How much impact do magnetic fields have on the
properties of the turbulence?  Crutcher (1999), using observations of
Zeeman splitting, found magnetic fields in some clouds strong enough
that one cannot safely neglect their effects.  Although it has been
shown that magnetic fields do not appreciably lengthen the turbulent
decay time (S98; Mac Low 1999), they do create anisotropy within the
clouds (e.g.~Vestuto et al.~2003, hereafter V03; Esquivel et
al.~2003), which may have important observational
and evolutionary consequences.  For example, molecular clouds are
often observed to be filamentary (e.g.~Mizuno et al.~1995; Churchwell
et al.~2004).

In this paper, we will investigate the energy dissipation properties
of supersonic hydrodynamic and MHD turbulence with a variety of
magnetic field strengths using Athena, a new higher-order Godunov
code.  An important goal of this study is to investigate the effect of
the assumed driving mechanism on the properties of the resulting
turbulence, such as power spectra and intermittency indicators.  Our
analysis utilizes data from high-resolution numerical simulations with
twenty-five different parameter sets.  In recent years, a variety of
results have been reported on the properties of supersonic MHD
turbulence, including energetics (e.g.~S98; Mac Low 1999; Ostriker et
al.~2001), power spectra (e.g.~Cho \& Lazarian 2003, 2005; V03;
Padoan et al.~2007, hereafter P07), and probability distribution
functions (e.g.~Padoan et al.~1997; Passot \& Vazquez-Semadeni 1998;
Ostriker et al.~2001; Kowal et al.~2007; Kritsuk et al.~2007,
hereafter K07). Where possible, we identify differences in our methods
and results as compared to those of other groups.  In a separate letter
(Lemaster \& Stone 2008, hereafter Paper I), we have reported the
results of an investigation of the variation of the probability
distribution function (PDF) of density with Mach number.  Our primary
result in that paper was that the intermittent behavior of turbulence
could be responsible for the large cloud-to-cloud variation in the
observed star formation rate per solar mass.

The primary tool available to investigate the properties of highly
nonlinear, supersonic MHD turbulence is direct numerical simulation.
To date, most results have been computed using a few numerical
algorithms, such as ZEUS (e.g.~S98; V03; Mac Low 1999; Ostriker et
al.~1999; Ostriker et al.~2001), the PENCIL code (e.g.~Haugen \&
Brandenburg 2004), the Stagger code (e.g.~P07), and ENO methods
(e.g.~Cho \& Lazarian 2003).  There has been concern expressed in
the literature that previous results may be strongly affected by
numerical dissipation.  Thus, it is worth re-examining the problem
with more accurate methods.  This study represents one of the first
applications of higher-order Godunov methods to the study of
supersonic MHD turbulence.

Without knowing the driving mechanism, we are left with many possible
methods of generating turbulence in simulations.  The hope is that the
choice of method for the simulated driving will have an insignificant
effect on the results.  Federrath, Klessen, \& Schmidt (in prep.),
however, have shown that compressive and solenoidal forcing produce
dramatically different turbulent states.
For some diagnostics, such as intermittency, the time-variability of
the turbulent state is critical.  Even if an array of driving methods
produce the same mean state, do the instantaneous states have the same
distribution about the mean?  We will investigate the dependence on
driving scale of various diagnostics, given our particular driving
method, which is very similar to that employed by, e.g., S98 and V03.

For power spectra of simulated turbulence to be valuable, the
resolution needs to be high enough that the driving and numerical
dissipation scales are well separated, allowing the inertial range to
be studied.  When magnetic fields are taken into account, simulating
turbulence at these resolutions can be prohibitively expensive.
Another goal of this study is to investigate whether properties of
the power spectra and other diagnostics are converged at the numerical
resolutions feasible at the moment, up to $1024^3$.

We begin in \S\ref{sec:methods} by describing our numerical methods in
detail.  We proceed, in \S\ref{sec:results}, to present our results.
\S\ref{sec:satresults} includes a convergence study and Mach number
scaling study of saturation amplitudes, \S\ref{sec:specresults}
includes a power spectrum analysis, \S\ref{sec:sonicresults} analyzes
the sonic scale in $1024^3$ runs, and \S\ref{sec:mittresults}
considers time-variability and temperature intermittency in the
turbulence.  Finally, we summarize our results in \S\ref{sec:concl}
and discuss the implications.

\section{Numerical Methods}\label{sec:methods}

The simulations we present were conducted up to a resolution of
$1024^3$ with the Athena code (Gardiner \& Stone 2005, 2008; Stone et
al.~2008; Stone \& Gardiner 2008) on a three-dimensional Cartesian
grid of length $L = 1$ with periodic boundary conditions.  Athena
utilizes a higher-order Godunov scheme which exactly conserves mass,
momentum, and magnetic flux.  We solve the equations of ideal
isothermal MHD,
\begin{equation}\label{eq:mhdeqn1}
\frac{\partial \rho}{\partial t} + {\bf \nabla} \cdot (\rho {\bf v}) = 0,
\end{equation}
\begin{equation}\label{eq:mhdeqn2}
\frac{\partial \rho {\bf v}}{\partial t} +
{\bf \nabla} \cdot (\rho {\bf vv} - { \bf BB} + P + B^2/2) = 0,
\end{equation}
and
\begin{equation}\label{eq:mhdeqn3}
\frac{\partial {\bf B}}{\partial t} =
{\bf \nabla} \times ({\bf v} \times {\bf B}),
\end{equation}
where $c_s = 1$ and $P = c_s^2 \rho$ are the isothermal sound speed
and pressure, respectively.  Because our method of driving the
turbulence, described in the following paragraph, does not involve
modifying the equations of MHD, we include no explicit forcing term
here.  We use an approximate nonlinear Riemann solver (HLLD; Miyoshi
\& Kusano 2005) for our MHD runs and an exact nonlinear Riemann solver
for our hydrodynamic runs.  Both our MHD and hydro simulations are
integrated well past the turbulent saturation time using a
directionally-unsplit van Leer scheme (Stone \& Gardiner 2008).
Further details of how we overcame the numerical difficulties
associated with utilizing this method for turbulence can be found in
Appendix \ref{sec:godunov}.

We initialize a uniform, stationary ambient medium with density
$\bar{\rho} = 1$ and magnetic field parallel to the $x$-axis whose
amplitude $B_0$ is fixed by the value of
$\beta = 2c_s^2\bar{\rho}/B_0^2$.  We then apply divergence-free
velocity perturbations following a Gaussian random distribution with a
Fourier power spectrum of the form
\begin{equation}
|\delta {\bf v}_k^2| \propto k^6 \exp (-8k/k_{\rm pk})
\end{equation}
for $kL/2\pi < N/2$, where $N$ is the resolution and $k_{\rm pk}$ is
the wavenumber of peak driving, in all but two of our runs.  For the
remaining two runs, with $\edot = 500$ and \pk{2} at $1024^3$, we
truncate the driving spectrum at $kL/2\pi = 8$.  Before applying the
perturbations to the grid, we shift them such that no net momentum
will be added and normalize them to give the desired energy injection
rate, \edot, or, in the decaying case, initial kinetic energy.

\begin{figure}
\epsscale{1.0}
\plotone{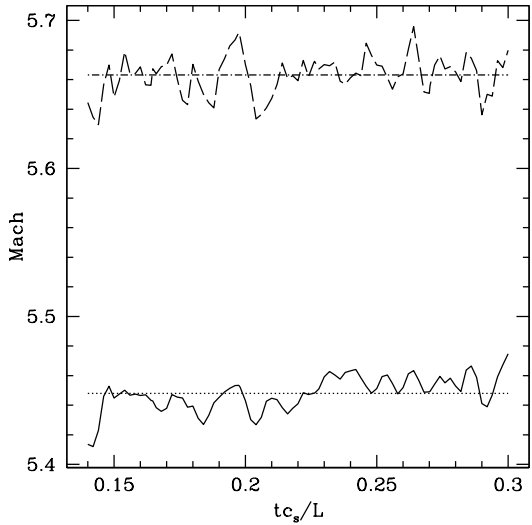}
\figcaption{
Comparison of turbulent Mach number computed with (solid) and without
(dashed) density-weighting for the driven, strong-field MHD turbulence
with \pk{8} and $\edot = 1000$.  The mean of the latter (dash-dotted)
is $\sim 4\%$ higher than that of the former (dotted).  With the
exception of for our sonic scale analysis, we use density-weighted
velocity dispersions.
\label{fig:machmeth}}
\end{figure}

For our driven runs, we choose our energy injection rate to give the
desired turbulent Mach number, $\mathcal{M} \equiv \sigma_v/c_s$, in
the saturated state.  Here, 
$\sigma_v = [\sigma_{v_x}^2+\sigma_{v_y}^2+\sigma_{v_z}^2]^{1/2}$ is
the 3D velocity dispersion of the gas and $\sigma{v_i}$ is the 1D
velocity dispersion in a given direction.  The method we use in this
paper to compute the turbulent Mach number differs from that used by
some other groups; we utilize a density-weighted velocity dispersion,
$\sigma_{v_i} = [\langle \rho v_i^2 \rangle/\langle \rho \rangle-
(\langle \rho v_i \rangle/\langle \rho \rangle)^2]^{1/2}$.  Note that,
due to the zero net momentum of our turbulent medium, the second term
on the RHS vanishes on the global scale.  If we compare the Mach
number computed with and without density-weighting the velocity
dispersion, we find that the two are fairly correlated, but that the
time-average of the latter is $\sim 4\%$ larger than that of the
former for driven strong-field MHD turbulence with, e.g., \pk{2} and
$\edot = 500$, or \pk{8} and $\edot = 1000$.  The comparison for the
latter case is shown in Figure \ref{fig:machmeth}.  We will use a
density-weighted velocity dispersion for all but our sonic scale
analysis.

To drive turbulence in our simulations, we inject energy before each
time step, with a new realization of the power spectrum generated at
intervals $\Delta tc_s/L = 0.001$.  This differs from that done in S98
and V03, where the energy was injected only when a new realization of
the spectrum was generated.  Ostriker et al.~(1999, 2001) and Kowal
et al.~(2007) used different driving spectra than our
own but still constrained their velocity perturbations to be
divergence free.  Our method differs from that of K07 in that they
approximate an isothermal equation
of state using an adiabatic index of $\gamma = 1.001$, drive their
turbulence on very large scales using an acceleration that is fixed in
time, and also allow a substantial fraction of the energy introduced
by the forcing to be in compressional modes.  P07 also used a fixed
acceleration for turbulent driving.  It is important to understand if
these differences have any significant impact on the results.

For our decaying runs, we again begin with a uniform, stationary
ambient medium but apply only one driving impulse to the velocity
field, with a power spectrum of the same form as is used to
initialize our driven runs.  We then allow it to evolve undisturbed.
We choose
the turbulent kinetic energy applied to our medium so as to give the
desired initial turbulent Mach number.  Our decaying runs differ from
those of Sytine et al.~(2000), who used a different initial driving
spectrum and allowed a compressible component in the initial
perturbations.  They used an adiabatic equation of state with
$\gamma = 1.4$, which we will use only for our decaying simulations.
In this case, we solve the total energy equation,
\begin{equation}\label{eq:mhdeqn4}
\frac{\partial E}{\partial t} + \nabla \cdot
\Big[(E+P+B^2/2){\bf v}-{\bf B}({\bf B}\cdot{\bf v})\Big] = 0,
\end{equation}
in addition to Equations (\ref{eq:mhdeqn1})--(\ref{eq:mhdeqn3}),
using the HLLC Riemann solver.  Here,
\begin{equation}
E = \frac{P}{\gamma-1}+\frac{1}{2}\rho v^2+\frac{1}{2}B^2
\end{equation}
with $P = (\gamma-1)e$ and $B^2 = {\bf B}\cdot{\bf B}$,
where $e$ is the internal energy density and $\gamma$ is the ratio of
specific heats.  The numerical methods in Athena conserve total energy
exactly.

We primarily investigate strong-field MHD turbulence with
$\beta = 0.02$ and hydrodynamic turbulence ($\beta = \infty$),
although we present runs with $\beta = 0.2$ (moderate-field) and
$\beta = 2.0$ (weak-field) as well.  Note that, due to a definition of
$\beta$ in S98 and V03 that differs from ours by a factor of 2, our
$\beta = 0.02$ results should be compared to their $\beta = 0.01$
results, and similarly for other values of $\beta$.
The magnetic fields we use in our simulations correspond to physical
values of $B = 2.0 \unit{\mu G} \beta^{-1/2} (T/10 \unit{K})^{1/2}
(n_{H_2}/10^2 \unit{cm^{-3}})^{1/2}$, where $T$ is the temperature and
$n_{H_2}$ is the number density of molecular hydrogen.  Our
simulations are scale-free, allowing them to be scaled to any set of
physical parameters using appropriate choices of $\bar{\rho}$, $c_s$,
and $L$.  Utilizing the same values given in S98, i.e.
$L = 2 \unit{pc}$, $n_{H_2} = 10^3 \unit{cm^{-3}}$, and
$T = 10 \unit{K}$, yields energy injection rates of up to
$\dot{E} = 0.4 L_\odot$, with a magnetic field strength
$B = 44 \unit{\mu G}$ for the strong-field case.


\section{Results}\label{sec:results}

\begin{figure*}
\epsscale{1.0}
\plotone{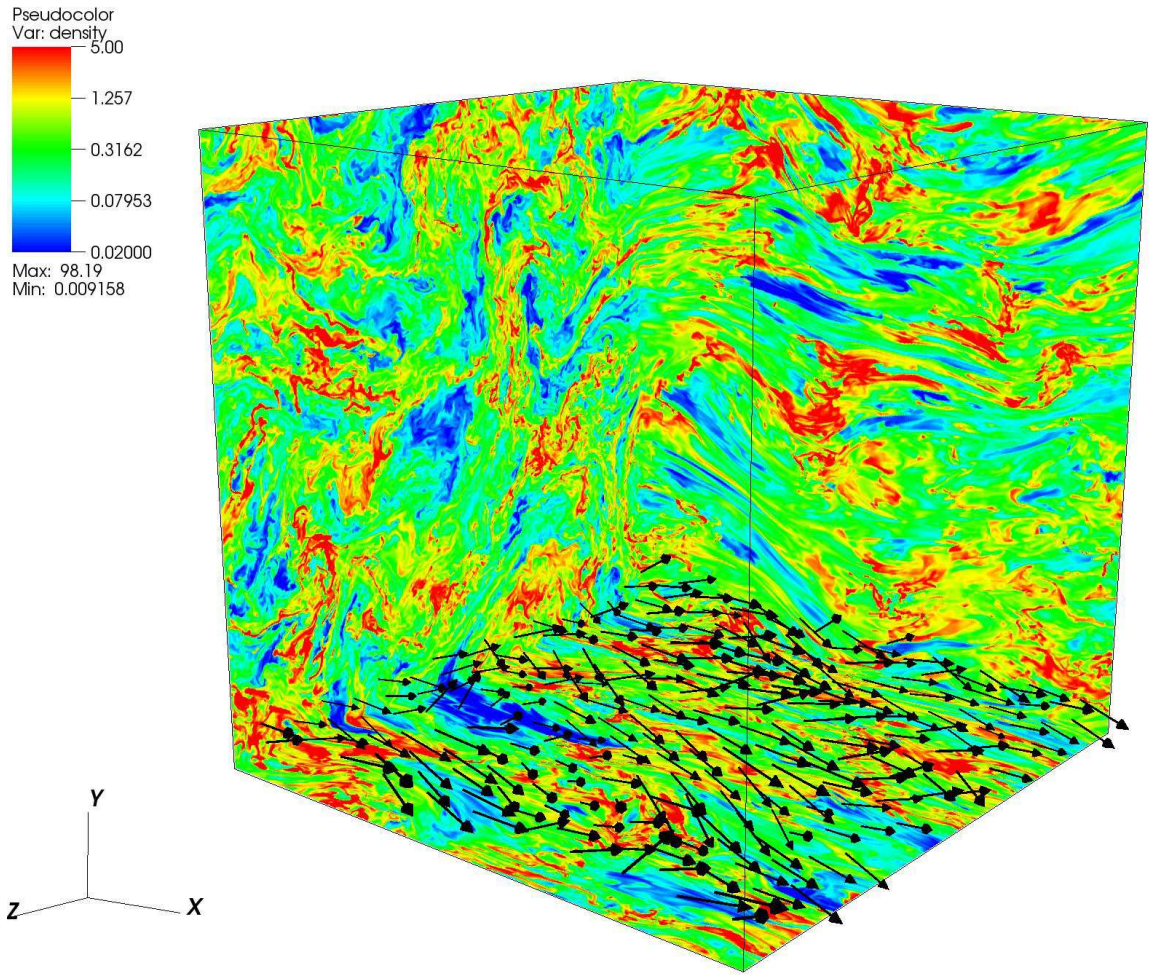}
\figcaption{
Driven strong-field $\mathcal{M} \approx 6.9$ MHD turbulence with
\pk{2} at $1024^3$.  Slices of density along the far faces of the cube
on a logarithmic color scale from 0.02 (blue) to 5.0 (red).  Magnetic
field vectors along a slice normal to the y-axis at $y \approx 0.0625$
are fairly well aligned and there is anisotropy in the scale of the
structures that results from the magnetic field.
\label{fig:den1024p2a}}
\end{figure*}

\begin{figure*}
\epsscale{1.0}
\plotone{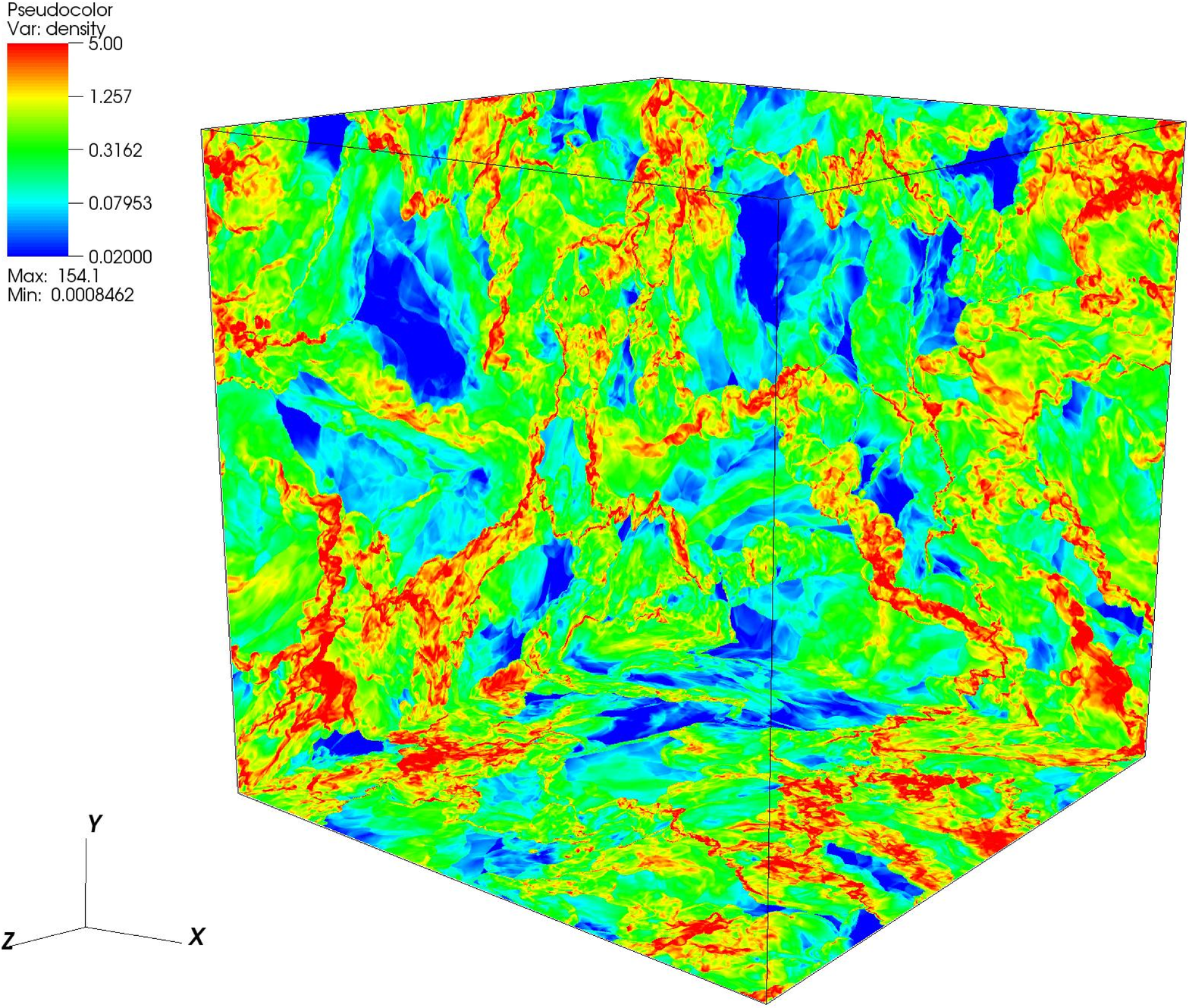}
\figcaption{
Driven $\mathcal{M} \approx 7.0$ hydrodynamic turbulence with \pk{2}
at $1024^3$.  Slices of density along the far faces of the cube on a
logarithmic color scale from 0.02 (blue) to 5.0 (red).  The structures
in this case are isotropic.
\label{fig:den1024p2d}}
\end{figure*}

\begin{figure*}
\epsscale{1.0}
\plotone{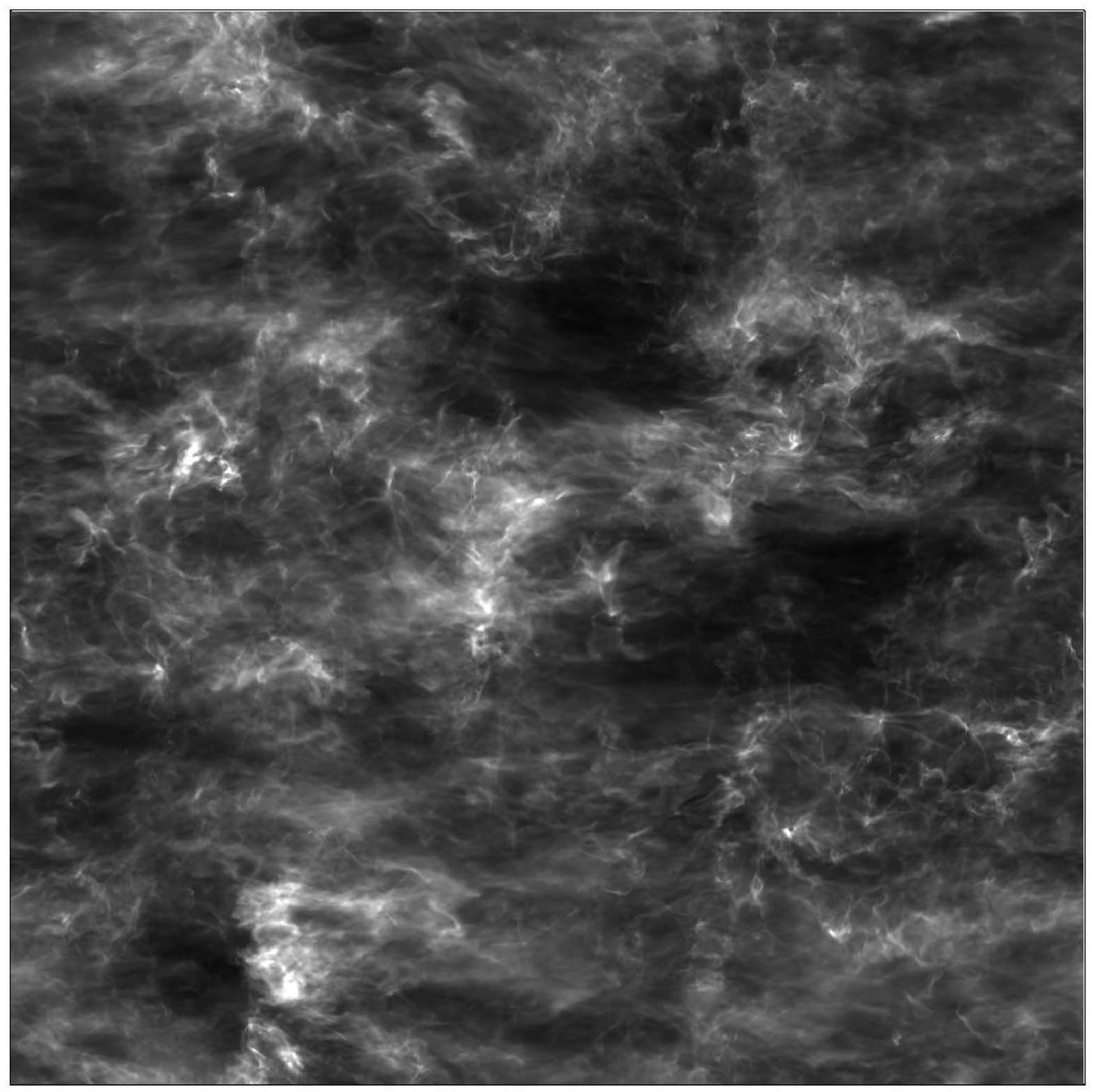}
\figcaption{
Column density along line of sight parallel to z-axis for driven
strong-field MHD turbulence with \pk{2} at $1024^3$ on a linear gray
scale from 0.3 (black) to 3.4 (white).
\label{fig:col1024p2a}}
\end{figure*}

\begin{figure*}
\epsscale{1.0}
\plotone{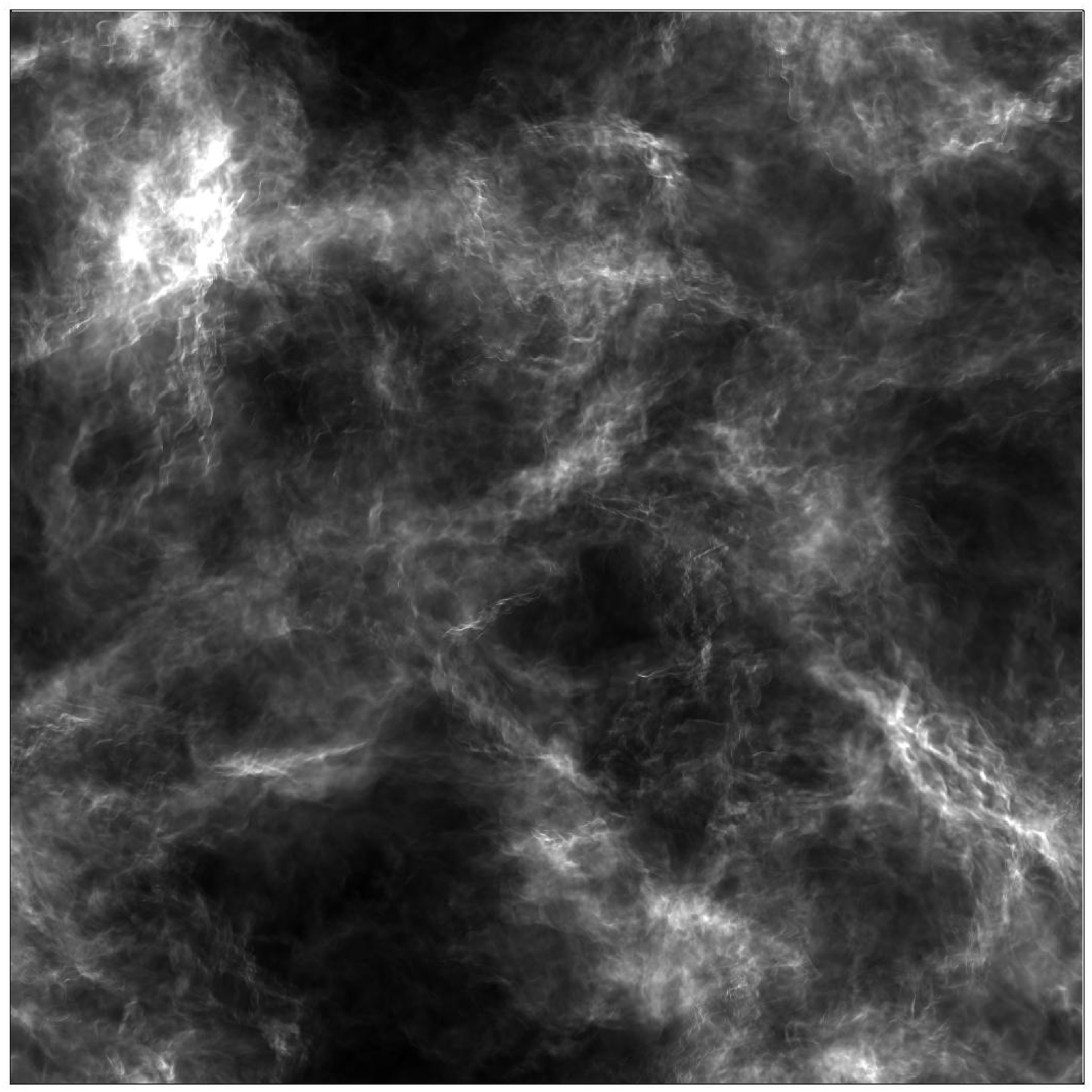}
\figcaption{
Column density along line of sight parallel to z-axis for driven
hydrodynamic turbulence with \pk{2} at $1024^3$ on a linear gray scale
from 0.3 (black) to 3.4 (white).
\label{fig:col1024p2d}}
\end{figure*}

Figure \ref{fig:den1024p2a} shows slices in mass density along the far
faces of the computational domain for driven $\mathcal{M} = 6.9$
strong-field MHD turbulence with \pk{2} at $1024^3$.  Also included are
magnetic field vectors in a slice normal to the y-axis near the bottom
of the cube.  Due to the strong background magnetic field, the vectors
are fairly well aligned.  Figure \ref{fig:den1024p2d} also shows slices
in density, but this time for $\mathcal{M} = 7.0$ hydrodynamic
turbulence with the same driving scale and resolution.  For both cases
the structure seen is filamentary, with anisotropy in the scale of the
structures in the strong-field case as a result of the field.  This is
not to say, however, that the filaments are aligned with the magnetic
field.  In fact, some appear to be oriented perpendicular to the
field.  Figures \ref{fig:col1024p2a} and \ref{fig:col1024p2d} show
column density along the line of sight parallel to the z-axis for the
same runs as in Figures \ref{fig:den1024p2a} and \ref{fig:den1024p2d},
respectively.  The projected filamentary structure is visible in these
column density images.  The column density contrast is higher for the
strong-field than for the hydrodynamic case.

\subsection{Saturation Amplitude}\label{sec:satresults}

We begin our quantitative analysis by studying the energy in
fluctuations once our driven turbulence runs have reached saturation.
Since the method with which we drive our turbulence injects energy at
a constant rate, at saturation the energy dissipation rate of the
turbulence will, on average, equal the energy injection rate.  At
sufficiently high numerical resolution, the numerical dissipation
will become negligible compared to the physically interesting sources
of dissipation (shocks), and the energy dissipation properties of the
turbulence can given us insight into the heating within and evolution
of molecular clouds.  In this section, we will first determine the
resolution at which the saturation energy has converged, i.e. reached
a state such that further increasing the resolution has a negligible
effect on the state, for turbulence evolved using Athena and compare
the turbulent decay rates to those presented in S98 computed using
ZEUS.

The kinetic energy associated with the fluctuations in our turbulent
medium can be quantified by $E_K \equiv \int \rho v^2/2 \, d{\bf x}$,
where $\rho$ is the density and $v$ is the magnitude of the velocity.
An integration for this and all similar quantities is performed over
the entire domain.  Similarly, the energy in magnetic field
perturbations is
$\delta E_B \equiv \int (B^2-B_0^2)/2 \, d{\bf x}$, where $B$ and
$B_0$ are the magnitudes, respectively, of the magnetic field and its
initial value.  The total energy in perturbations is the sum of
the energy in kinetic and magnetic field fluctuations,
$E_{\rm pert} \equiv E_K + \delta E_B$.  To analyze the energy
dissipation properties of the turbulence, we compute dissipation
timescales using $t_{\rm diss} = E/\dot{E}$, normalizing them to the
flow crossing time, $t_{\rm f} = \lambda_{\rm pk}/\sqrt{2E_K}$, where
$\lambda_{\rm pk} = 2\pi/k_{\rm pk}$ is the wavelength of peak
driving.

We can partition the kinetic energy by breaking down the velocity into
its solenoidal and compressive components.  The solenoidal component
is divergence-free, whereas the compressive component is curl-free.
These can easily be computed using
${\bf v}_C({\bf k}) = [\hat{k} \cdot {\bf v}({\bf k})] \hat{k}$ and
${\bf v}_S({\bf k}) = [\hat{k} \times {\bf v}({\bf k})] \times \hat{k}$,
respectively.
For simulations with comparable parameters (i.e.
\pk{8} and $\edot = 1000$),
these energies can be compared directly with the values given in
S98 and V03 after accounting for the difference in definition of
$\beta$.  We average our quantities over at least one dynamical time,
often several, beginning after the turbulence has fully saturated.

To determine the rate of convergence of a given volume-integrated
quantity, $q$, we compute the percent error in the quantity at each
resolution relative to the converged value.  In the case where the
quantity changes monotonically with resolution, we find the converged
value by performing a three-parameter fit,
$(q_N-q_\infty)/q_\infty = c N^{-\alpha}$, where $N$ is the
resolution and $q_N$ is the value at that resolution.  The results of
this fit tell us (1) the converged value, $q_\infty$, (2) the order of
convergence, $\alpha$, and (3) the coefficient, $c$, that determines
the resolution at which our result has converged, i.e. when
$(q_N-q_\infty)/q_\infty < 0.01$.

\subsubsection{Hydrodynamic Convergence Study}

To determine the resolution at which our numerical dissipation has
become small compared to shock dissipation, we
analyze the properties of driven, supersonic hydrodynamic turbulence
at resolutions from $32^3$ to $512^3$.  We study two sets of
simulations with identical energy injection rate, $\edot = 1000$, but
differing driving scales.  The set of runs with small-scale driving,
\pk{8}, correspond to the hydrodynamic run in S98, while the other set
of runs are driven at twice the scale, \pk{4}.  The
properties of the $512^3$ run from each set can be found in
Table \ref{tab:hydro}.  We have not performed a convergence study of
the \pk{2} runs due to the high level of time-variability (discussed
in \S\ref{sec:varresults}).

\begin{deluxetable*}{cccccccrrr}
\tablecolumns{10}
\tablewidth{0pc}
\tablecaption{Driven Hydro Turbulence at $512^3$\label{tab:hydro}}
\tablehead{
\colhead{$k_{\rm pk}L/2\pi$} 
& \colhead{$\edot$} 
& \colhead{$\mathcal{M}$} 
& \colhead{$E_K$} 
& \colhead{$E_C$}
& \colhead{$E_S$} 
& \colhead{$t_{\rm diss}/t_{\rm f}$}
& \colhead{$\sigma_E/E$} 
& \colhead{$\sigma_{Q,V}/Q_V$}
& \colhead{$\sigma_{Q,M}/Q_M$}}
\startdata
8 &  1000 &  5.8 &   17 &   3.8 &   13 &  0.77 & $<1\%$ &   --    &    --   \\
\hline
4 &  1000 &  7.2 &   26 &   5.9 &   20 &  0.76 & $ 1\%$ & $  1\%$ & $  2\%$ \\
4 &   375 &  5.3 &   14 &   3.1 &   11 &  0.78 & $ 1\%$ & $  1\%$ & $  2\%$ \\
4 &   140 &  3.8 &  7.2 &   1.6 &  5.6 &  0.78 & $ 1\%$ & $  2\%$ & $  2\%$ \\
4 &    40 &  2.6 &  3.3 &   0.7 &  2.6 &  0.84 & $ 1\%$ & $ <1\%$ & $ <1\%$ \\
4 &   3.5 &  1.2 &  0.8 &   0.1 &  0.7 &   1.1 & $<1\%$ & $ <1\%$ & $ <1\%$ \\
\hline
2 &   500 &  7.0 &   25 &   5.5 &   19 &   0.7 & $ 4\%$ & $  2\%$ & $  4\%$ \\
2 & 187.5 &  5.2 &   13 &   3.1 &   10 &  0.73 & $ 2\%$ & $  1\%$ & $  4\%$ \\
2 &    70 &  3.7 &  6.9 &   1.6 &  5.4 &  0.74 & $ 2\%$ & $  2\%$ & $  2\%$ \\
2 &    20 &  2.5 &  3.1 &   0.6 &  2.5 &  0.77 & $ 1\%$ & $  3\%$ & $  3\%$ \\
2 &  1.75 &  1.2 &  0.7 &   0.1 &  0.6 &  0.99 & $ 2\%$ & $  4\%$ & $  4\%$
\enddata
\end{deluxetable*}


Although both sets of runs are driven with energy injection rate
$\edot = 1000$, the larger driving scale of the \pk{4} set causes it
to converge to $\mathcal{M} \approx 7.2$, while the \pk{8} set
converges to only $\mathcal{M} \approx 5.8$.  The former set reaches
convergence by $64^3$, with higher resolutions having a small scatter
about the converged value.  The latter set, on the other hand,
converges monotonically at order $\alpha \approx 1.6$ for all
resolutions analyzed, reaching within $1\%$ of the converged value by
$128^3$.


The kinetic energy in fluctuations of the set with larger driving
scale, \pk{4}, converges to a value of $E_K \approx 26$ by $64^3$
and has some scatter about the converged value for higher
resolutions.  For the set with smaller driving scale, \pk{8}, the
kinetic energy in fluctuations converges at order
$\alpha \approx 1.5$ to only $E_K \approx 17$.  The value at $128^3$
is within $2\%$ of the converged value, while at $256^3$ it is within
a fraction of a percent.


Although we have an extremely low sampling rate for the fraction of
the kinetic energy in solenoidal and compressive modes, they appear to
be independent of driving scale.  The solenoidal fraction converges to
$E_S/E_K \approx 0.78$, decreasing with increasing resolution, while
the compressive fraction converges to $E_C/E_K \approx 0.22$.  The
solenoidal fraction is within a fraction of a percent of the converged
value by $256^3$; however, it varies by only a small amount down to
low resolution.  Finally, the ratio of the energy dissipation
timescale to the flow crossing time at the driving scale increases with
resolution, to $t_{\rm diss}/t_{\rm f} \approx 0.78$ for the \pk{4} set
and $t_{\rm diss}/t_{\rm f} \approx 0.76$ for the \pk{8} set, only a
small difference.  Consistent with our previous results, the \pk{4}
set converges at lower resolution than the \pk{8} set, $64^3$ and
$256^3$, respectively.  The turbulence dissipates in less than a
flow crossing time in all cases.


Although the converged Mach number and saturation energies are higher
for \pk{4} than for \pk{8}, the fraction of the energy in the
solenoidal or compressive mode, as well as the ratio of the energy
dissipation timescale to the flow crossing time at the driving scale,
are relatively independent of the driving scale.  While most of
the quantities of interest converge between $128^3$ and $256^3$ for
the \pk{8} hydrodynamic runs, convergence has already been reached for
these quantities by $64^3$ for the \pk{4} runs.  High resolutions are
critical for separating the driving and dissipation scales in the
power spectra far enough to study the inertial range; however,
quantities such as energy dissipation rate and turbulent Mach number
converge at resolutions which are more easily attainable.

\subsubsection{MHD Convergence Study}

We now analyze driven, strong-field, supersonic MHD turbulence in the
same manner as in the previous section.  To determine the resolution
at which our numerical dissipation becomes negligible, we study the
convergence of two sets of MHD simulations with energy injection rate
$\edot = 1000$ at resolutions from $32^3$ to $512^3$.  One set has
\pk{8}, similar to the strong-field MHD run in S98, while the other
has \pk{4}.  The properties of the $512^3$ run from each set
can be found in Table \ref{tab:mhd}.  Again, we have not performed a
convergence study of the \pk{2} runs due to the high level of
time-variability (discussed in \S\ref{sec:varresults}).

\begin{deluxetable*}{cccccccccrrr}
\tablecolumns{12}
\tablewidth{0pc}
\tablecaption{Driven MHD Turbulence at $512^3$\label{tab:mhd}}
\tablehead{
\colhead{$\beta$} 
& \colhead{$k_{\rm pk}L/2\pi$} 
& \colhead{$\edot$} 
& \colhead{$\mathcal{M}$} 
& \colhead{$E_K$} 
& \colhead{$E_C$}
& \colhead{$E_S$} 
& \colhead{$E_B$} 
& \colhead{$t_{\rm diss}/t_{\rm f}$}
& \colhead{$\sigma_E/E$} 
& \colhead{$\sigma_{Q,V}/Q_V$}
& \colhead{$\sigma_{Q,M}/Q_M$}}
\startdata
 2.0 & 8 &  1000 &  5.3 &   14 &   2.1 &   12 &  5.8 &  0.83 & $<1\%$ &   --    &   --    \\
 0.2 & 8 &  1000 &  5.1 &   13 &   1.6 &   11 &  8.6 &  0.89 & $<1\%$ &   --    &   --    \\
0.02 & 8 &  1000 &  5.4 &   15 &   1.5 &   13 &  8.2 &   1.0 & $<1\%$ &   --    &   --    \\
\hline
0.02 & 4 &  1000 &  6.8 &   23 &   2.3 &   21 &   13 &  0.98 & $<1\%$ & $  1\%$ & $  2\%$ \\
0.02 & 4 &   375 &  5.2 &   14 &   1.1 &   13 &  6.4 &   1.1 & $<1\%$ & $ <1\%$ & $  2\%$ \\
0.02 & 4 &   140 &  3.8 &  7.3 &   0.5 &  6.7 &  3.1 &   1.1 & $ 3\%$ & $  2\%$ & $ <1\%$ \\
0.02 & 4 &    40 &  2.7 &  3.6 &   0.2 &  3.4 &  1.4 &   1.3 & $<1\%$ & $  3\%$ & $  3\%$ \\
0.02 & 4 &   3.5 &  1.3 &  0.8 &  0.04 &  0.8 &  0.3 &   1.7 & $<1\%$ & $  2\%$ & $  2\%$ \\
\hline
0.02 & 2 &   500 &  7.0 &   24 &   2.2 &   22 &   11 &  0.98 & $ 4\%$ & $  4\%$ & $  5\%$ \\
0.02 & 2 & 187.5 &  5.0 &   13 &   1.2 &   11 &  5.9 &  0.99 & $ 3\%$ & $  2\%$ & $  2\%$ \\
0.02 & 2 &    70 &  3.7 &  6.8 &   0.6 &  6.3 &  2.9 &   1.0 & $ 3\%$ & $  4\%$ & $  5\%$ \\
0.02 & 2 &    20 &  2.5 &  3.1 &   0.2 &  2.9 &  1.3 &   1.1 & $ 2\%$ & $  3\%$ & $  4\%$ \\
0.02 & 2 &  1.75 &  1.2 &  0.8 &  0.06 &  0.7 &  0.4 &   1.6 & $ 6\%$ & $  9\%$ & $  8\%$
\enddata
\end{deluxetable*}

\begin{figure}
\epsscale{1.0}
\plotone{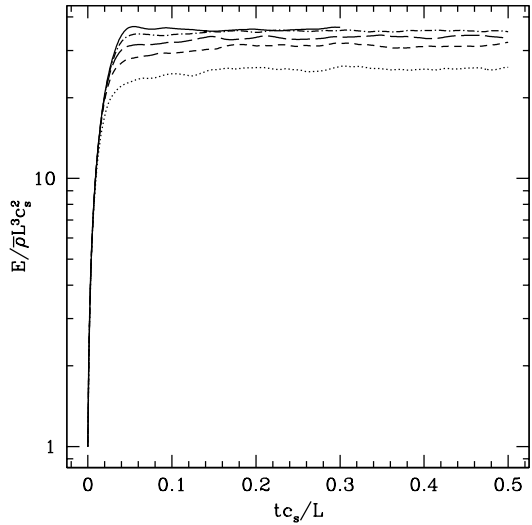}
\figcaption{
Total energy in fluctuations in driven MHD turbulence runs with
\pk{4} at $32^3$ (dotted), $64^3$ (short dashed),
$128^3$ (long dashed), $256^3$ (dash dotted), and $512^3$ (solid).
The saturated energy has converged by $256^3$.
\label{fig:mpert4res}}
\end{figure}

As in the hydrodynamic case, the \pk{4} MHD runs converge to a higher
Mach number, $\mathcal{M} \approx 6.8$, than the \pk{8} runs
($\mathcal{M} \approx 5.5$).  The former is converged by
$256^3$, while the latter converges at $1^{\rm st}$ order, only
reaching convergence at $512^3$.
The total energy in fluctuations converges at nearly $1^{\rm st}$
order to $E_{\rm pert} \approx 36$ for \pk{4}, shown in Figure
\ref{fig:mpert4res}.  While convergence has definitely been reached
by $512^3$, the $256^3$ value is less than $2\%$ from the converged
value.  For \pk{8}, on the other hand, convergence is not
reached until $512^3$, where $E_{\rm pert} \approx 23$.  Convergence
is at nearly $1^{\rm st}$ order for this case as well.

For the energy in magnetic field perturbations, convergence is reached
at $512^3$ for both the \pk{4} and \pk{8} cases.  Convergence is
approached at roughly 0.7 order for both cases.  The fraction of the
total energy (magnetic plus kinetic) in magnetic field fluctuations
increases with resolution, reaching $35\%$ at $512^3$.  When driven at
\pk{4}, the kinetic energy in fluctuations converges to
$E_K \approx 23$ by around $256^3$, while at \pk{8} it converges at
order 0.9 to $E_K \approx 15$ by $512^3$.


The fraction of the kinetic energy fluctuations in solenoidal modes
increases with resolution for both the \pk{4} and \pk{8} cases,
converging to $E_S/E_K \approx 0.9$.  The fraction in compressive
modes, on the other hand, decreases with resolution to
$E_C/E_K \approx 0.1$.  Convergence for these quantities is not
reached until $512^3$.  Finally, the ratio of the dissipation
timescale to the flow crossing time at the driving scale converges to
$t_{\rm diss}/t_{\rm f} \approx 0.98$ by $256^3$ for \pk{4}, while for
\pk{8} it converges to $t_{\rm diss}/t_{\rm f} \approx 1.0$ by $512^3$.
Even with a strong magnetic field, the turbulence dissipates in a flow
crossing time.


Just as in the hydrodynamic case, the increased driving scale caused a
higher turbulent Mach number and energy at saturation for the \pk{4}
than for the \pk{8} case.  The fraction of the kinetic energy in
solenoidal or compressive modes, however, was independent of driving
scale, just as before.  The ratio of the dissipation timescale to flow
crossing time
was also relatively independent of the driving scale.  Convergence for
strong-field MHD turbulence was reached at higher resolutions,
typically $512^3$, than for hydrodynamic turbulence.

\subsubsection{Comparison to ZEUS}


Because we performed a convergence study using the same turbulence
parameters as in S98, we can directly compare our high-resolution
results to those published therein.  For driven strong-field MHD
turbulence with \pk{8}, we find at $256^3$ that our
total energy in fluctuations at saturation is only $8\%$ higher than
that found in S98, due to the lower level of numerical dissipation at
this resolution in Athena than in ZEUS.  For intermediate- and
weak-field MHD, our energies are $7\%$ and $8\%$ higher, respectively.
It is likely that neither our results nor the S98 results have
converged by this resolution, however.

At $512^3$, there is no obvious difference between the total energy
in fluctuations for our strong-field MHD simulation and the S98
result.  The values from these two codes converge, even though they
utilize completely different numerical methods.  The ratio of magnetic
to kinetic energy fluctuations, however, is different for the two
codes, yielding ratios of dissipation timescale to flow crossing time
that differ more substantially.  At $256^3$, this ratio is $14\%$
greater for our strong-field MHD run than that presented in S98.  For
intermediate- and weak-field MHD, our ratios are both $11\%$ greater.
In all cases, the ratio of timescales remains below unity.


For our driven hydro run at $256^3$ with \pk{8}, we
find that our total energy in fluctuations at saturation is also only
$8\%$ higher than that found in S98.  Although our result has
converged by this resolution, the ZEUS result may not yet be
converged.  The ratio of dissipation timescale to flow crossing time
in this case is $11\%$ greater than that found in S98, but is still
below unity.  Although it has been suggested in the literature that
the rapid decay of supersonic turbulence is due to excessive numerical
dissipation in ZEUS, these results, computed with a higher-order
Godunov scheme, do not support that conclusion.

\subsubsection{Hydrodynamic Mach Number Scaling}
\label{sec:hydmachsatresults}

To investigate the effect of the turbulent Mach number on energies and
dissipation rates in the turbulence, we now analyze two series of five
driven, supersonic hydrodynamic turbulence simulations at $512^3$.
One series is driven at \pk{4}, while the other is driven at a larger
scale of \pk{2}.  The energy injection rates of the latter series,
the same that we use in Paper I, are half that of the
former series, giving roughly equal Mach numbers to the corresponding
pairs.  These Mach numbers all fall within the range
$1.2 \leq \mathcal{M} \leq 7.2$.  The properties of these runs can be
found in Table \ref{tab:hydro}.


The kinetic energy in each series of runs spans 1.5 orders of
magnitude.  We find a
power law relationship between the kinetic energy and energy injection
rate, i.e.
\begin{equation}
E_K \approx 0.49 [(k_{\rm pk}L/2\pi)/2]^{-1/2} (\edot)^{0.62}.
\end{equation}
The relationship between the kinetic energy and Mach number is, of
course, exactly $E_K = 0.5 \mathcal{M}^2$ since our Mach numbers are
computed from the {\em density-weighted} velocity dispersion and there
is no net momentum associated with the turbulent medium as a whole.
We find for our driving method that we can estimate the Mach number
that will result from a given energy injection rate using
\begin{equation}
\mathcal{M} \approx 0.99 [(k_{\rm pk}L/2\pi)/2]^{-1/4} (\edot)^{0.31}.
\end{equation}
While this equation gives an indication of how the dissipation scales
with Mach number, the exact relations will only apply to our unique
driving method.


For the runs driven at \pk{4}, the fraction of the kinetic energy in
compressive modes varies by only a fraction of a percent among the
runs with $3.8 \leq \mathcal{M} \leq 7.2$.  These runs have
$E_C/E_K \approx 0.22$.  The fraction drops off only slightly for the
$\mathcal{M} = 2.6$ run, but then drops substantially, to
$E_C/E_K \approx 0.14$, for the $\mathcal{M} = 1.2$ run.  For the runs
driven at \pk{2}, the fraction for the $\mathcal{M} = 7.0$ and
$\mathcal{M} = 3.7$ runs is also $E_C/E_K \approx 0.22$, while the
$\mathcal{M} = 5.2$ and $\mathcal{M} = 2.5$ runs have a bit larger and
smaller fractions, respectively.  Once again, the fraction for the
$\mathcal{M} = 1.2$ run is substantially smaller,
$E_C/E_K \approx 0.15$.  It appears as though the fraction of the
energy in compressive modes remains roughly the same except when the
turbulence is only mildly supersonic, where the fraction is much lower.


The ratio of the dissipation timescale to the flow crossing time at
the driving scale remains below unity for all except the
$\mathcal{M} = 1.2$ run at \pk{4}.  While the ratio for the runs at
\pk{4} with $\mathcal{M} \geq 3.8$ doesn't vary much, when we
increase the driving scale to \pk{2} the value does change with
Mach number.  The ratio for the \pk{4} runs is always larger for a
given Mach number than is the ratio for the \pk{2} runs.

\subsubsection{MHD Mach Number Scaling}

Finally, we analyze two series of five driven, supersonic,
strong-field MHD turbulence simulations at $512^3$ to determine the
effect of the turbulent Mach number on the energies and dissipation
rates.  As before, one series is the same \pk{2} set of runs used in
Paper I, while the other is driven at \pk{4} with
twice the energy injection rate, yielding pairs of simulations of
roughly the same Mach number within the range
$1.2 \leq \mathcal{M} \leq 7.0$.  The properties of these runs can be
found in Table \ref{tab:mhd}.


As in the hydro case, the kinetic energy in each strong-field MHD
series spans nearly 1.5 orders of magnitude.  The
total energy in fluctuations, which also includes the energy in
magnetic field perturbations, increases by an even larger amount from
the lowest to highest Mach number runs.  The power law relationship
between the Mach number and energy injection rate is
\begin{equation}
\mathcal{M} \approx
(1.04\pm 0.02) [(k_{\rm pk}L/2\pi)/2]^{-1/4} (\edot)^{0.30},
\end{equation}
which leads to a slightly narrower range of Mach numbers than does the
relation for pure hydro.  Again, although this gives an indication of
how energy dissipation scales with Mach number, the exact relationship
is unique to our driving method.


The fraction of the kinetic energy in compressive modes for
strong-field MHD is much less than for hydro.  When \pk{2}, the
fractions increase monotonically with Mach number, from only
$E_C/E_K \approx 0.05$ for $\mathcal{M} = 1.3$ to
$E_C/E_K \approx 0.10$ for $\mathcal{M} = 6.8$.  When \pk{4},
however, the values are much less predictable.  In this case, the
$\mathcal{M} = 3.7$ and $\mathcal{M} = 1.2$ runs have nearly
identical values when \pk{2}, whereas for hydro the fraction at
low Mach number was substantially smaller than the values for
higher Mach numbers.


The ratio of the dissipation timescale to the flow crossing time at
the driving scale decreases with Mach number.  While the values for
high Mach number are below unity, the low Mach number values become
as large as $t_{\rm diss}/t_{\rm f} \approx 1.7$.  The values for the
highest Mach numbers are roughly the same for both driving scales, but
for the smaller driving scale they increase more quickly.  The
Mach number has a strong influence on the dissipation rate of the
turbulence, but even for $\mathcal{M} \approx 1.2$ the dissipation
timescale does not exceed twice the flow crossing time.

\subsection{Power Spectra}\label{sec:specresults}


We next consider turbulent velocity power spectra.  We compute the
power spectrum (PS) of the velocity,
$P_K({\bf k}) = |{\bf v}({\bf k})|^2/2$,
as well as that of its compressive and solenoidal components,
$P_C({\bf k}) = |\hat{k} \cdot {\bf v}({\bf k})|^2/2$ and
$P_S({\bf k}) = |\hat{k} \times {\bf v}({\bf k})|^2/2$, respectively,
in the same way as was done in V03 for the PSD of the specific kinetic
energy and its components.  To generate the spherically-integrated
compensated power spectra that we will present, we average
$P({\bf k})$ within spherical shells and multiply by the volume within
the shell, $dV = 4\pi k^2 dk$, to find $P(k)$, where
$k = (k_x^2+k_y^2+k_z^2)^{1/2}$.

To determine the degree of anisotropy in MHD turbulence, we will also
analyze cylindrically-integrated power spectra.  We generate these
spectra by averaging $P_K({\bf k})$ over cylindrical shells whose
axes are oriented along the mean magnetic field direction.  This yields
$P_K(k_\parallel,k_\perp)$, where $k_\parallel = |k_x|$ is parallel to
the mean magnetic field and $k_\perp = (k_y^2+k_z^2)^{1/2}$ is in
the plane normal to the field.

The majority of the power spectra we present in our figures are
compensated---divided by a power law to produce plots where the
inertial range is very roughly horizontal,
making small deviations from a power law easier to see.  For each of
our driven turbulence runs, we compute power spectra from many
snapshots taken at regular intervals, averaging them together to
minimize the effects of rare events.  With the exception of the
$1024^3$ runs, we average over at least 69 snapshots spanning at
least 4.6 dynamical times.  While we averaged over 2.7 dynamical
times for the $1024^3$ hydro run, the MHD case is not sufficiently
time-averaged.

\subsubsection{Decaying Subsonic Hydrodynamic Turbulence}
\label{sec:specdecayresults}


Sytine et al.~(2000) presented the power spectra of decaying,
subsonic, adiabatic hydrodynamic turbulence at a range of resolutions.
These power spectra demonstrate the formation of a feature known as
the bottleneck effect.  Energy cascades down from larger scales but
cannot easily be dissipated, causing a build-up of small-scale power.
We begin by verifying that Athena can reproduce the bottleneck.  These
are the only set of adiabatic runs we will consider.

\begin{figure}
\epsscale{1.0}
\plotone{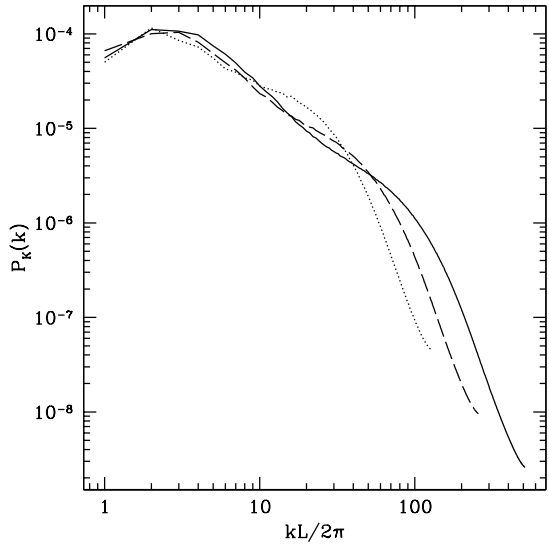}
\figcaption{
Spherically-integrated velocity power spectra at $tc_s/L = 2$ for
decaying  initially Mach 0.5 adiabatic hydro turbulence with
\pk{4}, at $1024^3$ (solid), $512^3$ (long dashed),
and $256^3$ (dotted).  Compare to Figure 11 of Sytine et al.~(2000).
\label{fig:hdcpwr1k4res}}
\end{figure}

The initial velocity perturbations in these decaying runs have the
same form of power spectrum as in our driven runs with \pk{4} and are
normalized to give an initial turbulent Mach number of
$\mathcal{M}_0 = 0.5$.  After this impulse is given to the initially
uniform ambient medium, it is allowed to evolve undisturbed until
$tc_s/L = 2$.  Figure \ref{fig:hdcpwr1k4res} shows the uncompensated
velocity power spectrum at this time for runs with resolutions from
$256^3$ to $1024^3$.  As expected, this plot appears very similar to
Figure 11 of Sytine et al.~(2000).  Since Athena was designed to have
low numerical dissipation, the bottleneck is strong in our simulations
of mildly-compressible (subsonic) turbulence.

Analyzing the compressive and solenoidal components of the velocity
separately, we find that the bottleneck appears quite strong in the
PS of the latter while apparently absent in that of the former.  This
is not surprising as shocks directly couple large and small scales,
allowing the compressive energy to bypass the turbulent cascade and be
immediately dissipated.  The absence of the bottleneck in the
compressive component agrees with the results of Porter et al.~(1999)
for Mach 1 driven, adiabatic hydro turbulence.  At the very highest
wavenumbers in our simulations, the power in
the compressive component flattens out.  Although this could be an
artifact of the code related to its treatment of shocks, one should be
wary of putting too much stock in the high-$k$ region of any power
spectrum as effects such as aliasing (introduced in the calculation of
the power spectrum, not the fluid dynamics) could substantially alter
the power in that region (see, e.g., Press et al.~1992 for a more
detailed discussion).

\subsubsection{Driven Hydrodynamic Turbulence}
\label{sec:spechydresults}




Next we consider driven supersonic (isothermal) hydrodynamic
turbulence.  We compute velocity power spectra for runs with \pk{2}
to maximize the separation between the driving and dissipative scales.
To obtain a slightly higher turbulent Mach number,
$\mathcal{M} \approx 7.0$, than before, we use an energy injection
rate of $\edot = 500$.  These are the same runs from the Mach
number scaling analysis of \S\ref{sec:hydmachsatresults}.

\begin{figure}
\epsscale{1.0}
\plotone{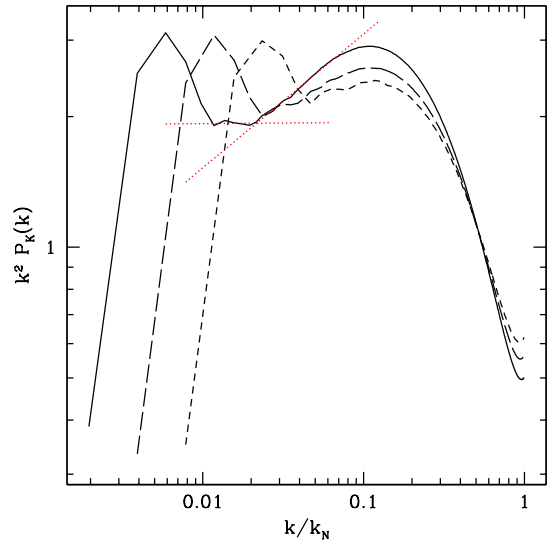}
\figcaption{
Spherically-integrated, compensated power spectra of velocity for driven
hydro turbulence runs with \pk{2} at $1024^3$ (solid), $512^3$ (long
dashed), and $256^3$ (short dashed).  The x-axis has been renormalized
to give $k/k_N = 1$.  Also shown are fits (dotted) to the slope of the
$1024^3$ run over the ranges $6 \le kL/2\pi \le 11$ and
$11 \le kL/2\pi \le 36$, representing power laws $P(k) \propto k^{-2.00}$
and $P(k) \propto k^{-1.69}$, respectively.
\label{fig:hpwr1k2res}}
\end{figure}

Figure \ref{fig:hpwr1k2res} shows our compensated time-averaged
velocity power spectrum for resolutions of $256^3$ through $1024^3$.
To align the dissipative range for simulations with different
resolutions, we express wavenumber as a fraction of the Nyquist value,
$k_NL/2\pi = N/2$ for a simulation with resolution $N^3$.
At low resolution we see no inertial range, but by $1024^3$ we have
separated the driving and dissipative scales enough that we may be
just starting to see an inertial range.  There is a small range,
$6 \le kL/2\pi \le 11$, where we see roughly a $P(k) \propto k^{-2}$
power law, but we are still using a small amount of forcing up
through $kL/2\pi = 8$ so the slope in this range may be affected.
We find a much longer power law range over $11 \le kL/2\pi \le 36$,
$P(k) \propto k^{-1.7}$.  Although it may be argued that a slope this
shallow must be due to a bottleneck, the range of scales with steeper
slope is far too limited to conclude this definitively from our data.
A higher resolution would be required to determine if this is actually
the case.

\begin{figure}
\epsscale{1.0}
\plotone{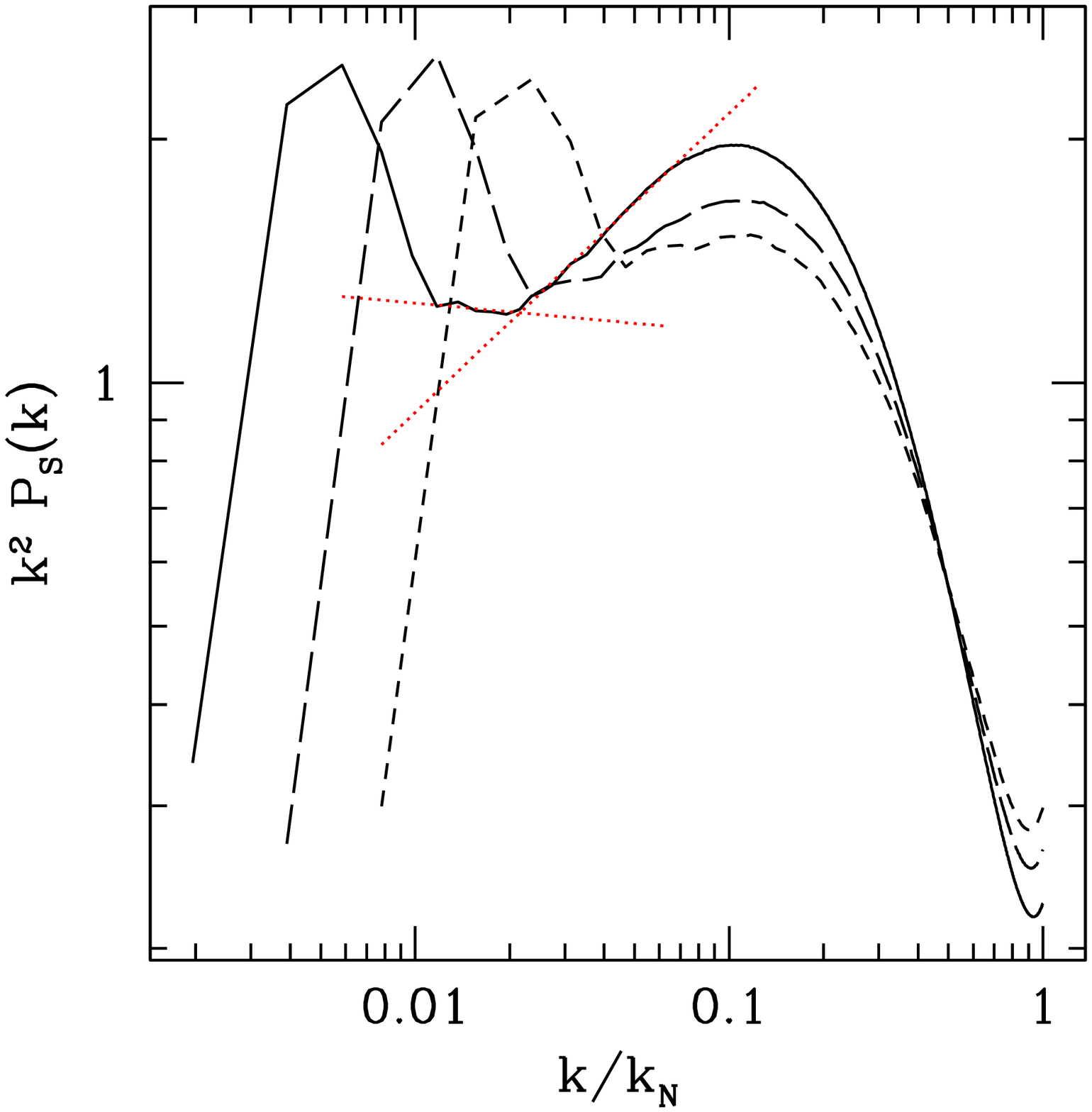}
\figcaption{
Spherically-integrated, compensated power spectra of the solenoidal
component of velocity for driven hydro turbulence runs with \pk{2} at
$1024^3$ (solid), $512^3$ (long dashed), and $256^3$ (short dashed).
The x-axis has been renormalized to give $k/k_N = 1$.  Also shown are
fits (dotted) to the slope of the $1024^3$ run over the ranges
$6 \le kL/2\pi \le 11$ and $11 \le kL/2\pi \le 36$, representing power
laws $P(k) \propto k^{-2.04}$ and $P(k) \propto k^{-1.63}$,
respectively.
\label{fig:hpwr1s2res}}
\end{figure}

\begin{figure}
\epsscale{1.0}
\plotone{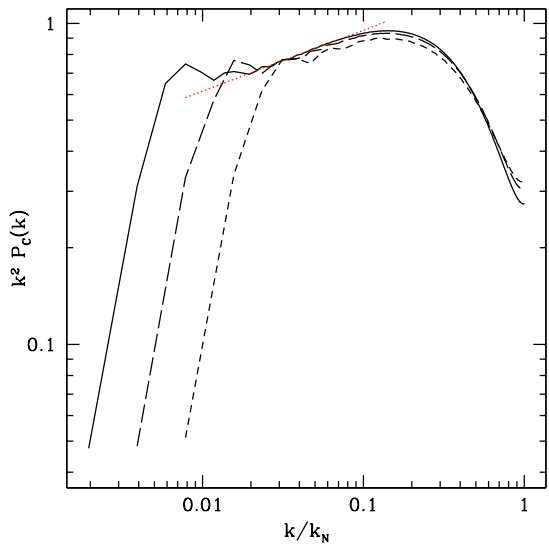}
\figcaption{
Spherically-integrated, compensated power spectra of the compressive
component of velocity for driven hydro turbulence runs with \pk{2} at
$1024^3$ (solid), $512^3$ (long dashed), and $256^3$ (short dashed).
The x-axis has been renormalized to give $k/k_N = 1$.  Also shown is a
fit (dotted) to the slope of the $1024^3$ run over the range
$10 \le kL/2\pi \le 40$, representing the power law
$P(k) \propto k^{-1.81}$.
\label{fig:hpwr1c2res}}
\end{figure}

Figures \ref{fig:hpwr1s2res} and \ref{fig:hpwr1c2res} show power
spectra of the solenoidal and compressive components of velocity,
respectively.  For the solenoidal component, we find power laws
$P(k) \propto k^{-2.0}$ and $P(k) \propto k^{-1.6}$ over the ranges
$6 \le kL/2\pi \le 11$ and $11 \le kL/2\pi \le 36$, respectively.
For the compressive component, however, we find only
$P(k) \propto k^{-1.8}$ over the range $10 \le kL/2\pi \le 40$.
In two independent $512^3$ simulations that use different random
perturbations to seed and drive the turbulence, we find the slopes
of the compressive component to agree with each other as well as
that of the $1024^3$ simulation to within a percent,
suggesting that the length of our time-averaging is sufficient.
The ratio of power in the compressive and solenoidal components of
velocity is much higher than for subsonic turbulence, having the
effect of slightly washing out the shallow feature in the velocity
spectrum in the supersonic case.

When comparing hydro turbulence with \pk{8}, we find more power at
high wavenumbers for turbulence evolved using Athena than using ZEUS.
Although it was stated in V03 that the bottleneck would not affect
finite difference codes, plotting the compensated power spectrum shows
some evidence of blending of the driving peak with a secondary bump or
shallowing of the spectrum (much less prominent than in our own).  If
this is indeed a bottleneck, this difference likely results from the
greater numerical dissipation in ZEUS than in Athena.

\subsubsection{Driven MHD Turbulence}\label{sec:specmhdresults}





\begin{figure}
\epsscale{1.0}
\plotone{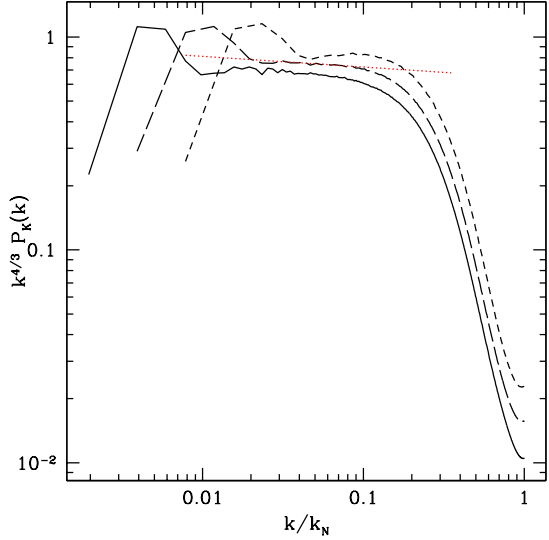}
\figcaption{
Spherically-integrated, compensated power spectra of the velocity for
driven strong-field MHD turbulence runs with \pk{2} at $1024^3$
(solid), $512^3$ (long dashed), and $256^3$ (short dashed).  The
x-axis has been renormalized to give $k/k_N = 1$.  Also shown is a fit
(dotted) to the slope of the $512^3$ run over the range
$8 \le kL/2\pi \le 18$, $P(k) \propto k^{-1.38}$.  The $1024^3$ run
appears only slightly steeper.
\label{fig:mpwr1k2res}}
\end{figure}
\begin{figure}
\epsscale{1.0}
\plotone{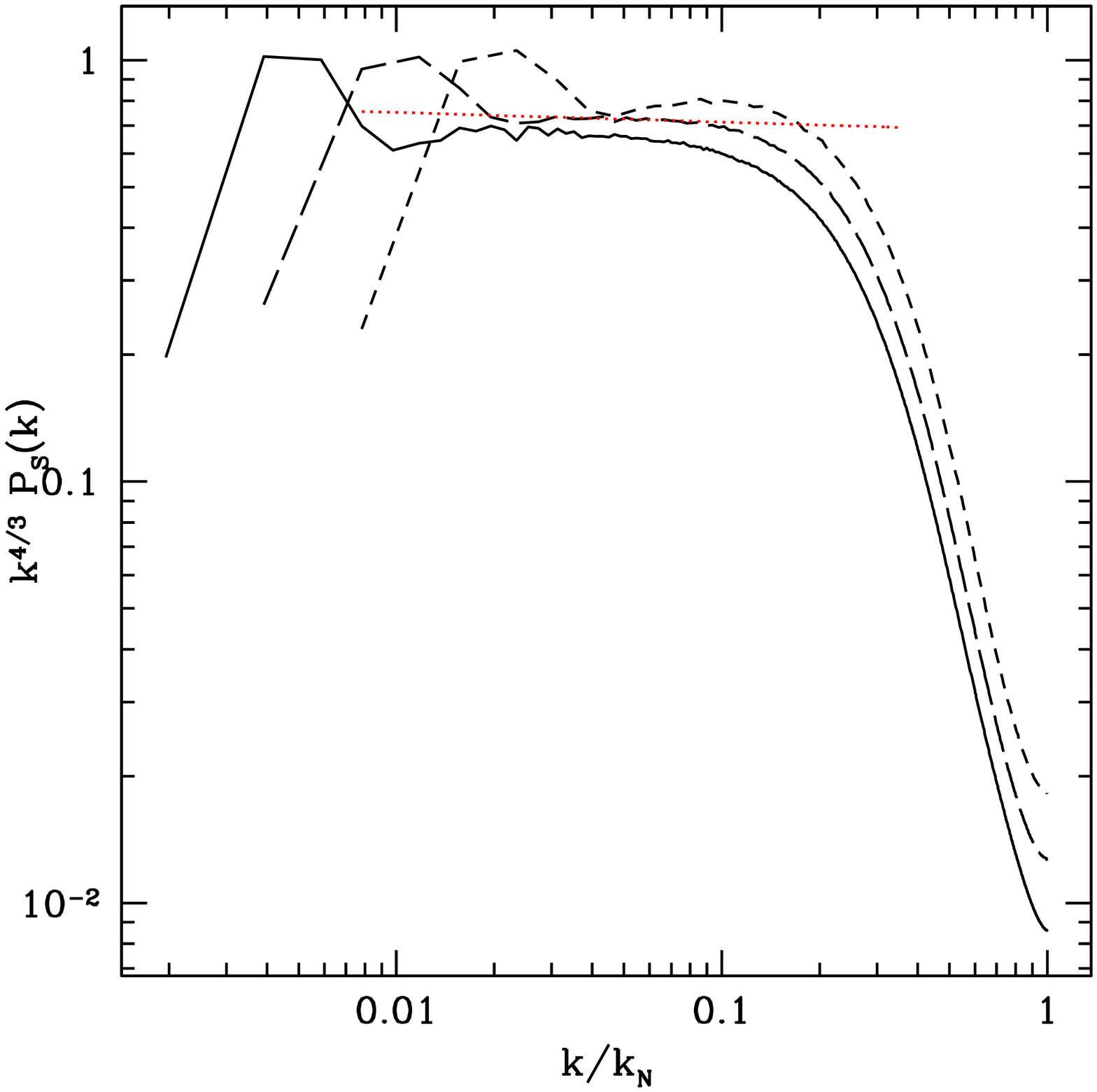}
\figcaption{
Spherically-integrated, compensated power spectra of the solenoidal
component of velocity for driven strong-field MHD turbulence runs with
\pk{2} at $1024^3$ (solid), $512^3$ (long dashed), and $256^3$ (short
dashed).  The x-axis has been renormalized to give $k/k_N = 1$.  Also
shown is a fit (dotted) to the slope of the $512^3$ run over the range
$8 \le kL/2\pi \le 18$, $P(k) \propto k^{-1.36}$.  The $1024^3$ run
appears only slightly steeper.
\label{fig:mpwr1s2res}}
\end{figure}

Analyzing driven strong-field MHD turbulence with \pk{2} and an energy
injection rate of $\edot = 500$, shown in Figure \ref{fig:mpwr1k2res},
we again find no inertial range at low resolution.  In this
$\mathcal{M} \approx 7.0$ run, the velocity power
spectrum is dominated by the solenoidal component, resulting in a
substantial shallowing of the spectrum that is apparent in the
higher-resolution runs.  At $512^3$, the slope of the spectrum over
the interval $8 \leq kL/2\pi \leq 18$ is only slightly steeper than
$P(k) \propto k^{-4/3}$ for the velocity and its solenoidal (Figure
\ref{fig:mpwr1s2res}) component.  We find spectral slopes in two
independent $512^3$ simulations (whose power spectra were averaged
together to give that plotted in the figures) that agree to within
a few percent, suggesting that our time-averaging to obtain a
reliable spectral slope.

Although we do not have a
sufficiently long time-average to find a robust fit to the slope of
the $1024^3$ run, the slopes appear to be very similar.  As before,
the slope we have found is much shallower than is typically predicted
predicted for the inertial range, while being steeper than that
presented in K07 as the bottleneck spectral slope.
We note that in the MHD case, the range over which we find a power
law seems to increase with resolution, which one would not
expect if the feature was due to numerical dissipation (bottleneck).

\begin{figure}
\epsscale{1.0}
\plotone{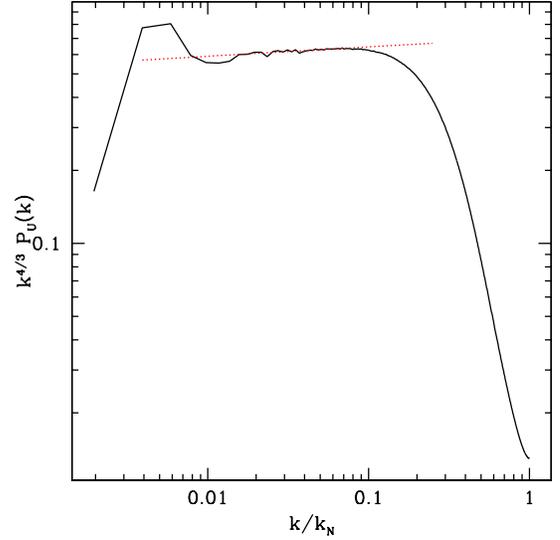}
\figcaption{
Spherically-integrated, compensated power spectra of the quantity
$u = \rho^{1/3} v$ for driven strong-field MHD turbulence with \pk{2}
at $1024^3$ (solid).  The x-axis has been renormalized to give
$k/k_N = 1$.  Also shown is a fit (dotted) to the slope over the range
$8 \le kL/2\pi \le 36$, $P(k) \propto k^{-1.29}$.
\label{fig:mpwr1u2res}}
\end{figure}

It has been argued by K07 that the quantity
${\bf u} = \rho^{1/3} {\bf v}$ should have an inertial range power
law of $k^{-5/3}$ even for supersonic turbulence, but we see over a
wide range of wavenumbers a $k^{-4/3}$ law (Figure \ref{fig:mpwr1u2res}).
Kurien et al.~(2004) have shown for subsonic turbulence that, while the
$k^{-5/3}$ law is appropriate when the energy timescale dominates over
the helicity timescale, a $k^{-4/3}$ law is expected when the helicity
timescale is non-negligible compared to the energy timescale, even
when the relative helicity may seem insignificant.  This should be
investigated as a possible cause of the bottleneck effect in
supersonic turbulence.

\begin{figure}
\epsscale{1.0}
\plotone{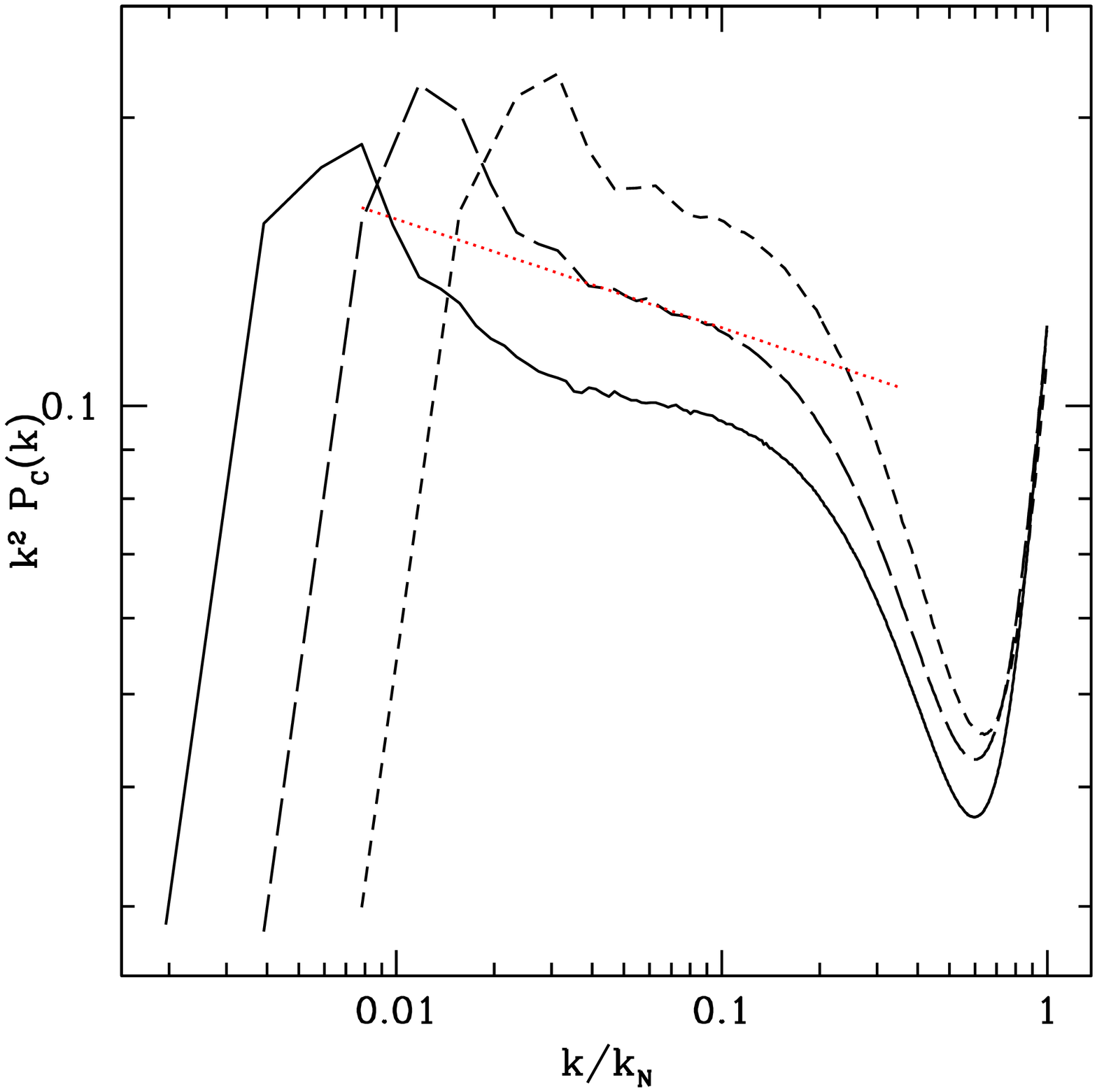}
\figcaption{
Spherically-integrated, compensated power spectra of the compressive
component of velocity for driven strong-field MHD turbulence runs with
\pk{2} at $1024^3$ (solid), $512^3$ (long dashed), and $256^3$ (short
dashed).  The x-axis has been renormalized to give $k/k_N = 1$.  Also
shown is a fit (dotted) to the slope of the $512^3$ run over the range
$10 \le kL/2\pi \le 18$, $P(k) \propto k^{-2.11}$.  The $1024^3$ run,
which does not have a sufficiently long time-average to be robust,
does not appear to have power law form.  The prominent upturn at the
highest wavenumbers is due to the uncompensated power spectrum
flattening out to a constant value.
\label{fig:mpwr1c2res}}
\end{figure}

Over the range $10 \le kL/2\pi \le 18$, the compressive component
(Figure \ref{fig:mpwr1c2res}) of the $512^3$ run follows a power law
$P(k) \propto k^{-2.1}$.  For $1024^3$ run, the power spectrum
appears to smoothly change slope, indicating that it is not a power
law, but a longer time-average might change the shape.  The power
spectrum of the magnetic field at $512^3$ follows the power law
$P(k) \propto k^{-1.2}$ over the range $8 \le kL/2\pi \le 18$.
At $1024^3$, the slope appears very similar, but with extra noise at
the smaller wavenumbers due to the short time-average used.  For all
resolutions studied, the uncompensated power approaches a constant,
non-zero value at the highest wavenumbers.  This produces a very
prominent upturn in the compensated spectra.

\begin{figure}
\epsscale{1.0}
\plotone{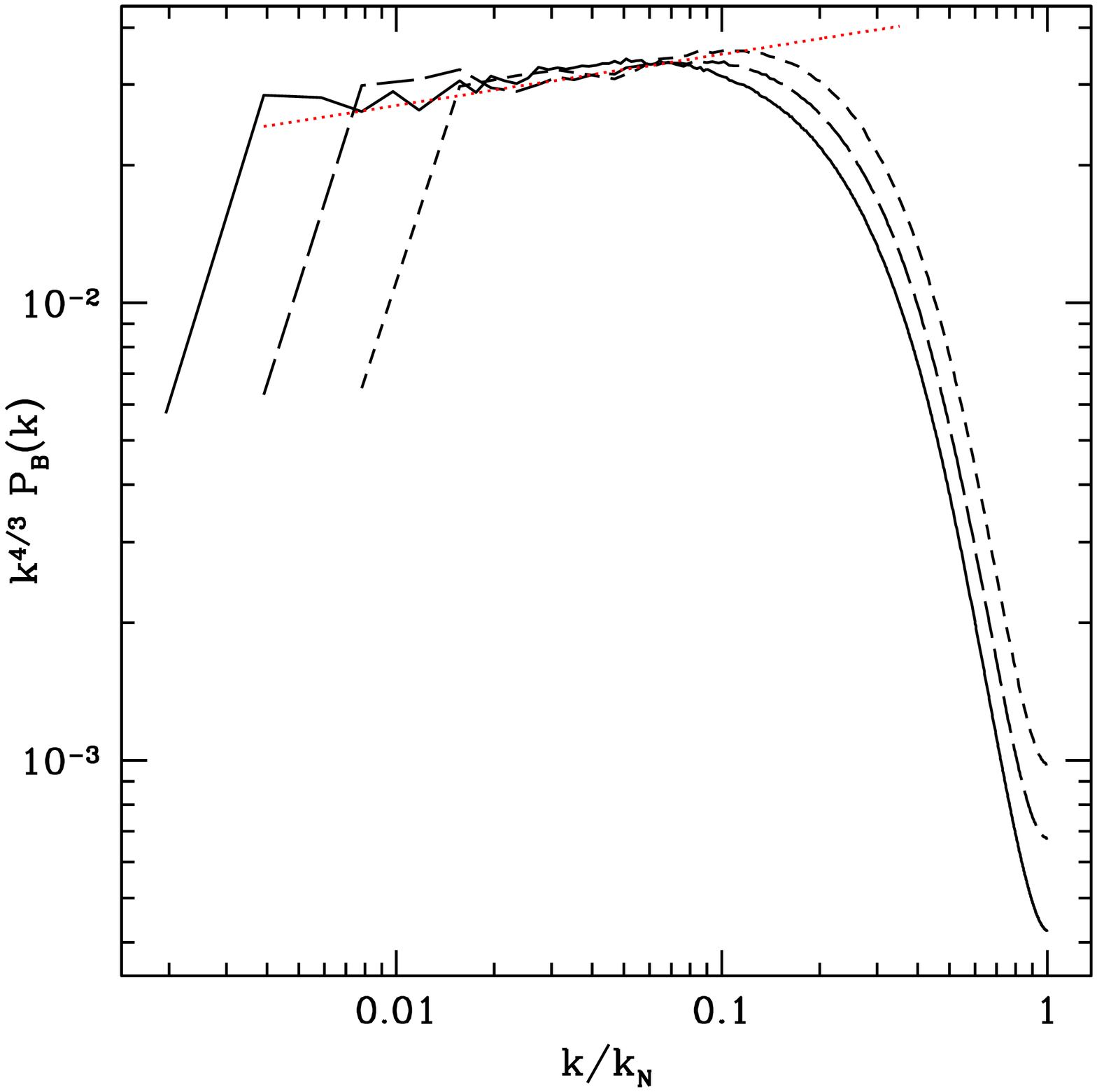}
\figcaption{
Spherically-integrated, compensated power spectra of the magnetic
field perturbations for driven strong-field MHD turbulence runs with
\pk{2} at $1024^3$ (solid), $512^3$ (long dashed), and $256^3$ (short
dashed).  The x-axis has been renormalized to give $k/k_N = 1$.  Also
shown is a fit (dotted) to the slope of the $512^3$ run over the range
$8 \le kL/2\pi \le 18$, $P(k) \propto k^{-1.22}$.
\label{fig:mpwr1b2res}}
\end{figure}

A direct comparison for strong-field MHD with \pk{8} shows
more power at high wavenumber from Athena than from ZEUS for the
velocity and its compressive and solenoidal components, just as was
true for the hydro case.  If the shallowing of our velocity spectrum
is due to the bottleneck effect, then what appeared to be inertial
range when V03 was published is likely affected by the bottleneck
as well.

\begin{figure}
\epsscale{1.0}
\plotone{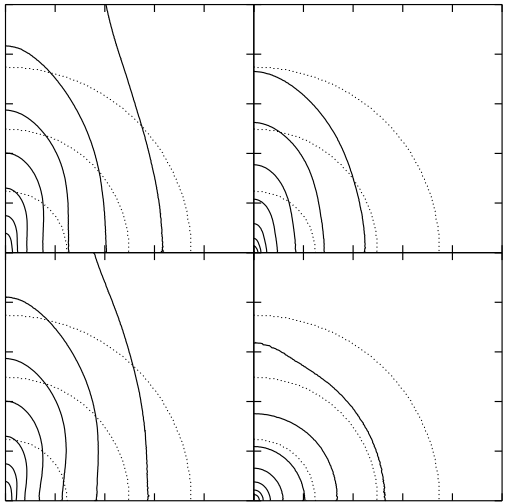}
\figcaption{
Cylindrically-integrated ({\em un}compensated) power spectra (solid)
of (a) the velocity, (b) the magnetic field, and the (c) solenoidal
and (d) compressive components of velocity for the driven strong-field
MHD turbulence run at $512^3$ shown in Figure \ref{fig:mpwr1k2res}.
The $x$ and $y$ axes are $k_\parallel/k_N$ and $k_\perp/k_N$,
respectively.  The colors correspond to $P_K(k_\parallel,k_\perp)$,
where $k^2 = k_\parallel^2+k_\perp^2$.  Also shown for reference are
circular contours (dotted).
\label{fig:mpwrspec2d}}
\end{figure}

Figure \ref{fig:mpwrspec2d} shows our cylindrically-integrated power
spectra at $512^3$.  We find anisotropy in the power spectrum of
magnetic field perturbations as well as the velocity and its
solenoidal component.  There is more power perpendicular to the mean
magnetic field at a given wavenumber than parallel to it.  Parallel to
the mean field, the power law is roughly $k^{-2}$, similar to that of
the purely hydrodynamic case.  Perpendicular to the field, however,
the slope is much more shallow, roughly $k^{-4/3}$.  We find the power
spectrum of the compressive component of velocity to be nearly
isotropic, in contradiction to what was found in V03.

\begin{figure}
\epsscale{1.0}
\plotone{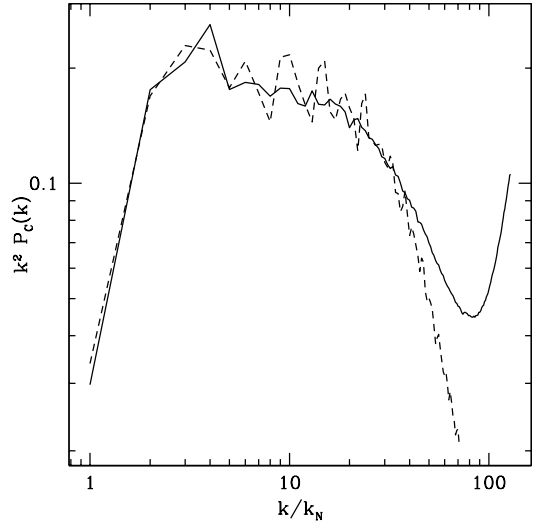}
\figcaption{
Spherically-integrated compensated power spectrum of the compressive
component of velocity from one snapshot of the $256^3$ driven
strong-field MHD turbulence run (solid) shown in Figure
\ref{fig:mpwr1k2res}, compared to the total velocity power spectrum of
an initially-spherical MHD blast wave (short dashed), also at $256^3$.
Except for the oscillations, the shapes of these spectra look quite
similar between the driving and dissipative scales.
\label{fig:blastcomprimh}}
\end{figure}

The velocity power spectrum of {\em subsonic} turbulence results from
a conservative cascade of energy from large to small scales through
interactions local in Fourier space.  It is commonly believed that
the velocity spectrum of supersonic turbulence, because it too is a
power law, must also result from such a cascade.  It is commonly
known, however, that the power spectrum of a discontinuity, or an
ensemble of shocks as in Burgers turbulence, also have velocity
spectra that follow a power law, $P(k) \propto k^{-2}$.  In Figure
\ref{fig:blastcomprimh}, we confirm that the compressive component
of velocity in our $256^3$ supersonic, strong-field MHD turbulence
run with \pk{2} has a spectrum similar to that of velocity in an MHD
blast wave that was initially spherical.  This calls into question
the long-held assumption that the power spectrum seen in supersonic
turbulence results from an energy cascade facilitated by interactions
local in Fourier space.  It would seem that other diagnostics, such
as structure functions, are necessary in order to determine the
cause of the power law spectrum, either a Fourier-space cascade as
in incompressible turbulence, or an ensemble of shocks as in Burgers
turbulence.

To get a clearer idea of which features in our power spectra are
representative of the turbulence and which were introduced in the
post-processing, we compare power spectra computed by multiple
methods.   Our control spectrum will be computed as were the others
presented in this section---averaged over the cells within a shell and
then multiplied by the volume of the shell, $dV = 4\pi k^2 dk$, which
we will refer to as ``shell-averaged'' for the remainder of this
section.  Our comparison will be to another commonly used method
(e.g.~P07), where the power is simply summed over the cells falling
within the shell (``shell-summed'').  These methods differ due to the
discretization of the grid, i.e. because the Fourier-space volume
occupied by the cells within a bin is not, in general, equal to the
volume of the perfectly spherical shell that the bin represents.
While the bins in our control spectrum are centered on integer
values of $kL/2\pi$, those in the alternative spectrum instead run
{\em between} integer values.

\begin{figure}
\epsscale{1.0}
\plotone{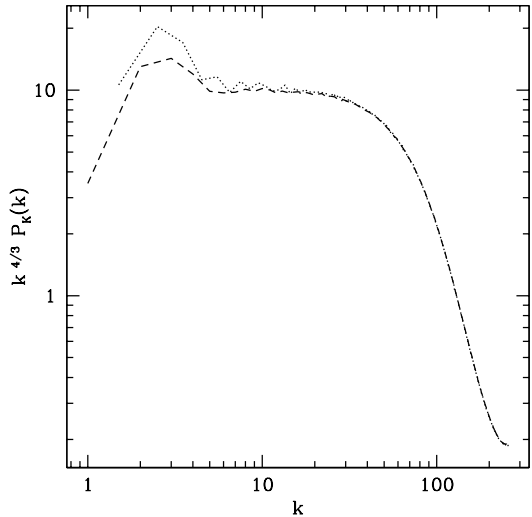}
\figcaption{
Spherically-integrated compensated power spectrum of the total
velocity for the $512^3$ driven strong-field MHD turbulence run shown
in Figure \ref{fig:mpwr1k2res} calculated using our standard method
(solid) compared to the method described in P07 (dotted).  The PS
computed using the P07 method is more jagged in the range
$5 < kL/2\pi < 20$.  Note that the compensation used for this pair of
power spectra is non-standard.
\label{fig:psmethod}}
\end{figure}

We use the time-averaged power spectrum of the driven strong-field MHD
turbulence run at $512^3$ for our comparison, although the effect
should be independent of the run being analyzed.  When overplotting
the spectra produced by these two methods (see Figure
\ref{fig:psmethod}), the most obvious difference is at low $k$.
Although the shell-averaged spectrum is relatively smooth in the range
$5 < kL/2\pi < 20$, the other has a jagged shape over this same range
resulting from its sensitivity to the number of cells falling within a
bin, making the ``slope'' of the power spectrum in this region much less
obvious.

If we compare the shell-summed spectrum to one computed by the same
method but with the bins shifted by half a bin width (aligning these
bins with our own), we find that shape of the spectrum at low $k$
changes considerably.  If, on the other hand, we take the
shell-averaged spectrum and compare it to one computed in the same
manner but with bins shifted to align with those typically used in the
shell-summed method, we see that, although the power in each bin does
change, the shape and slope of the spectrum change very little.
While neither method is right or wrong, we advise caution when
choosing a binning method for power spectra; differences in slope
computed by different means are not necessarily indicative of
different turbulent states.

\subsection{Sonic Scale}\label{sec:sonicresults}

Although molecular clouds have supersonic velocity dispersions on
large scales, as one looks to smaller and smaller scales, turbulent
compressions will become progressively weaker, at some point becoming
subsonic.  The length scale at which the RMS velocity dispersion is
equal to the sound speed is referred to as the sonic scale (McKee \&
Ostriker 2007).  We now investigate how the velocity dispersion varies
on spatial scales larger than the sonic scale, where shocks are most
important.  This scaling can be determined observationally in the
form of linewidth-size relations using many different methods.  For
example, Brunt (2003) used principal component analysis (PCA) to
determine the linewidth-size relation within individual clouds.

To determine the velocity dispersion at a given length scale,
$kL/2\pi = m$, where $m = 2^n$ and $n$ is an integer, we divide our
computational domain
along each axis into $m$ sections, yielding $m^3$ sub-volumes.  We
compute the velocity dispersion in each sub-volume and then average
over all sub-volumes at that scale.  Because crossing the sonic scale
represents a change in the physics dominating the flow, i.e. waves
steeping to form shocks, for this analysis only we will compute the
velocity dispersion without mass-weighting, i.e.
$\sigma_v = [\sigma_{v_x}^2+\sigma_{v_y}^2+\sigma_{v_z}^2]^{1/2}$,
where
$\sigma_{v_i} = [\langle v_i^2 \rangle-\langle v_i \rangle^2]^{1/2}$.

\begin{figure}
\epsscale{1.0}
\plotone{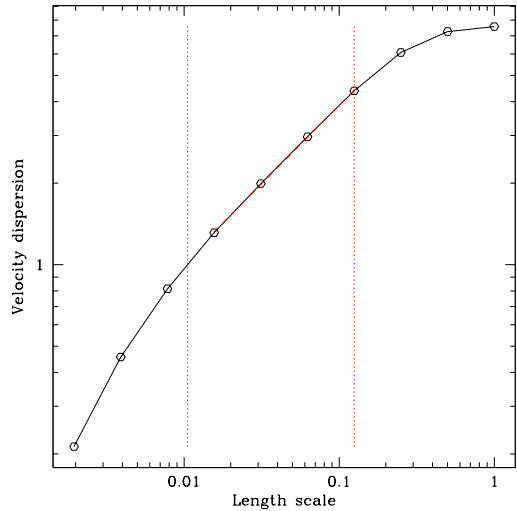}
\figcaption{
3D velocity dispersion versus the spatial scale on which it is
measured for driven hydrodynamic turbulence with \pk{2} (open
hexagons).  Also shown is a power law fit,
$\sigma{l} \propto l^{0.58}$, from the driving
cutoff ($kL/2\pi = 8$) down to our last data point above the sonic
scale (both limits marked with dashed lines).  We have connected our
data points with a solid line to make comparing to the fit easier.
\label{fig:sonichyd}}
\end{figure}

Figure \ref{fig:sonichyd} shows the velocity dispersion versus length
scale for the $1024^3$, $\mathcal{M} \sim 7$ driven hydrodynamic
turbulence run with \pk{2}.  Since driving may affect the scaling
relation, we truncate our driving spectrum at $kL/2\pi = 8$, where the
driving has already dropped to only $1\%$ of the peak level, and
consider only this and smaller scales.  Fitting to
the data points falling between the sonic scale and the driving
cutoff, a factor of eight in length, we find a clear power law of the
form $v(l) \propto l^{0.58}$.  This scaling index falls well within
the range of indices found observationally by Brunt (2003).

\begin{figure}
\epsscale{1.0}
\plotone{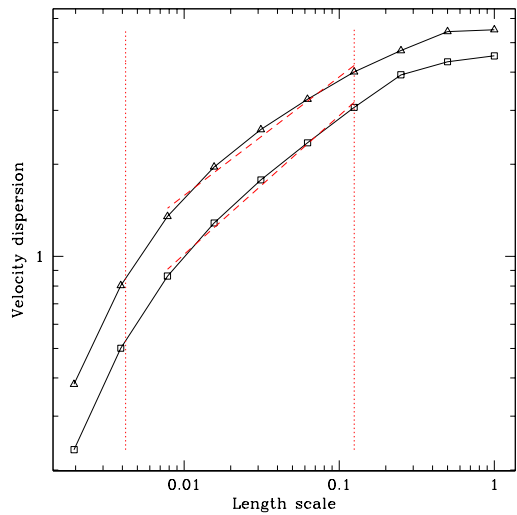}
\figcaption{
1D velocity dispersion parallel to the magnetic field (open squares)
and 2D velocity dispersion perpendicular to the field (open triangles)
versus the spatial scale on which it is measured for driven
strong-field MHD turbulence with \pk{2}.  From the driving cutoff
($kL/2\pi = 8$) down to our last data point before the 3D velocity
dispersion becomes subsonic (both marked with dashed lines), neither
the parallel nor perpendicular velocity dispersion has power law form
(dotted).  We have connected our data points with a solid line to make
comparing to the power laws easier.
\label{fig:sonicmhd}}
\end{figure}

Shown in Figure \ref{fig:sonicmhd} is the velocity dispersion versus
length scale for the $1024^3$, $\mathcal{M} \sim 7$ driven
strong-field MHD turbulence run with \pk{2}.  This time, over the
range of scales where the flow is supersonic, we find that the slope
is not a power law.  Unlike in hydrodynamic turbulence, the sonic
scale is not the only scale at which the dominant physics changes.
In fact, we would expect the multiple wave families of MHD to lead to
multiple transitions, and for strong fields, these transitions are
widely separated.  For example, parallel to the magnetic field, slow
waves will travel at the sound speed, whereas fast waves will travel
at the Alfv\'{e}n speed.  Perpendicular to the field, however, the slow
waves will have zero velocity and the fast waves will travel with a
speed $\sqrt{v_A^2+c_s^2}$.  This is complicated further by the fact
that the parallel and perpendicular directions are defined relative to
the {\em local} magnetic field, not the mean field.  As a result, we
have no reason to expect a uniform power law between the sonic scale
and driving cutoff.  It is not clear to what extent this result is
affected by dissipation.

\subsection{Time-variability and Temperature Intermittency}
\label{sec:mittresults}

Up to this point we have averaged quantities from multiple snapshots
in order to minimize the effect of rare events on our results.  Our
final set of diagnostics will instead be an analysis of the
time-variability and temperature intermittency of the turbulent gas.

\subsubsection{Saturation Amplitude}\label{sec:varresults}



We begin by analyzing the time-variability of some of the quantities
we analyzed in \S\ref{sec:satresults}.  For our hydrodynamic runs with
the smallest driving scale, \pk{8}, we find the time-variability of
the saturation energy to be only half a percent at both $256^3$ and
$512^3$.  Time-variability of less than a percent agrees with the
results in S98 even though they drove their turbulence impulsively.
The time-variability quoted in K07, however, is much higher.  The K07
runs were driven at a much larger driving scale than our own.  To
investigate if this could be the reason for the discrepancy, we will
also analyze the time-variability of our runs with larger driving
scales.

When we increase the driving scale to \pk{4}, we find the
time-variability of the energy to be a full percent at both $256^3$
and $512^3$.  As both the driving scale and Mach number are varied
here as compared to the run we discussed above, this could be due to
either.  When we increase the driving scale further, to \pk{2},
without changing the Mach number, we find the time-variability
increases yet again, coming to $2\%$ at $256^3$ and approaching $4\%$
at $512^3$.  It would seem that the driving scale chosen has a
significant impact on the time-variability of the energy.  The values
we find here are still less than that shown in the K07 figure,
however.


For strong-field MHD, we again find time-variability of about half a
percent for the \pk{8} runs.  Increasing the driving
scale to \pk{4} decreases the time-variability at
$256^3$ but increases it a small amount at $512^3$.  Increasing the
driving scale further, to \pk{2}, yields
time-variability of $3\%$ at $256^3$ and $4\%$ at $512^3$, comparable
to the hydro runs.

When we compare the time-variability of the energy among hydro runs at
$512^3$ with varied Mach number that are driven at
\pk{4}, we find that the relation is non-monotonic.
With \pk{2}, however, we do see a monotonic
relationship, with the time-variability increasing with Mach number.
For our strong-field MHD runs, we again find non-monotonicity at
\pk{4}, while the time-variability increases with
Mach number at \pk{2}, with the exception of the
lowest Mach number run.

\subsubsection{Temperature Intermittency}\label{sec:mitttemp}

A feature of turbulence that may have a strong influence on star
formation is intermittency, dissipation that is highly localized in
space and time (McKee \& Ostriker 2007).  Although such dissipation
does not require shocks, it is worthwhile to study the shock
contribution in supersonic turbulence.  Were we not assuming an
isothermal equation of state, energy dissipation in these regions
would lead to local heating and thus increased temperatures, which
would be evident in observations.  Therefore, to study intermittency
due to shocks, we analyze the maximum heating rate per unit mass,
$Q = c_s^2 \nabla \cdot {\rm v}$, a surrogate for temperature.

Since we expect the maximum heating rate, $Q_{\rm max}$, to be
strongly influenced by rare events, such as the interaction of
multiple shocks, we consider the time-variability of this quantity.
Due to the discretization of the grid, however, we would expect a
component of the time-variability to be due to grid-scale
fluctuations.  Therefore, instead of considering $Q_{\rm max}$
directly, we consider the statistics of the high-$Q$ tail of the
PDF of the heating rate per unit mass.  For each simulation that we
consider, we compute the heating rate per unit mass, $Q$, in each cell
over many snapshots.  Considering only the cells in each snapshot that
compose the $1\%$ of material (by mass or volume) with the highest
cell-averaged $Q$, we compute the mean value.  Following the
time-variability of this value should allow us to ignore meaningless
grid-scale fluctuations while studying the intermittency due to
shocks.


For driven hydrodynamic turbulence at $512^3$, we find the low Mach
number runs driven at \pk{4} to have less than $1\%$ variability in
the peak temperature, while the higher Mach number runs with this same
driving scale can have variability as large as $2\%$ but without a
monotonic dependence on Mach number.  At the larger driving scale
\pk{2}, however, the variability in the peak temperature always
exceeds $1\%$ and can be as high as $4\%$.  It would seem that this,
like the time-variability of the saturation energies, increases with
driving scale.  For our runs, the variability of the peak temperature
as measured by the top $1\%$ of the volume never exceeds that from
the top $1\%$ of the mass by more than a tiny amount.  As we have very
few data dumps with which to calculate the statistics, however, the
quantitative behavior may not be robust.

\begin{figure}
\epsscale{1.0}
\plotone{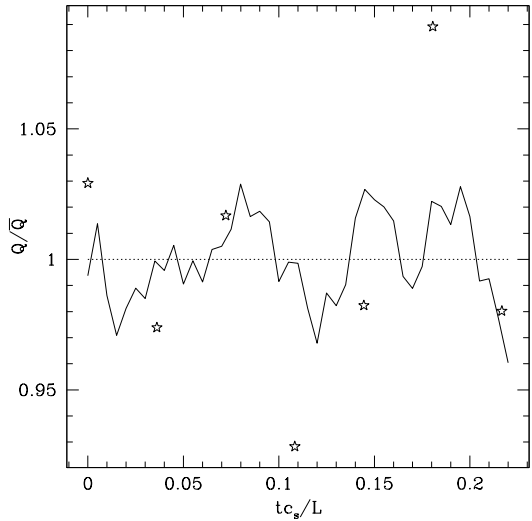}
\figcaption{
Maximum heating rate per unit mass (solid), where the tail (see
\S\ref{sec:mitttemp}) is defined by the top $1\%$ of mass, for driven
strong-field $\mathcal{M} \sim 7$ MHD turbulence with \pk{4} at
$512^3$.  The time-variability here is lower compared to the case
where \pk{2} (shown by stars) at roughly the same Mach number.
\label{fig:mmitt4}}
\end{figure}

We find that, in $512^3$ strong-field MHD turbulence driven at \pk{2},
the variability of the peak temperature can be as large as $9\%$.
When driven at \pk{4}, however, the variability always decreases,
never exceeding $3\%$.  The variability measured by the top $1\%$ of
mass is typically larger than that from the top $1\%$ of volume, but
there does not appear to be a strong dependence on Mach number.  More
often than not, the variability in the strong-field MHD runs are larger
than in the hydro runs at the same Mach number.  Again, however, these
statistics are computed from very few data dumps, making them subject
to large errors.  Figure \ref{fig:mmitt4} shows the time-evolution of
the peak temperature in the \pk{4} strong-field MHD run with
$\edot = 1000$, for which we have better than typical statistics,
compared to the \pk{2} run with $\edot = 500$.

\section{Discussion and Conclusions}\label{sec:concl}


The saturation energies and dissipation timescales we find support
the conclusion of S98 that supersonic turbulence dissipates rapidly
even in the presence of magnetic fields.  The results of our Godunov
code agree with ZEUS, an operator-split method that relies on
artificial viscosity to capture shocks, but reach convergence at
slightly lower resolutions.  At $512^3$ the difference in the total
energy in fluctuations between Athena and ZEUS for strong-field MHD
is very small, indicating that these two codes converge to similar
turbulent states at sufficiently high resolution.

The convergence resolution of our simulations depends both on the
driving scale and the presence of a magnetic field.  While hydro
turbulence driven at \pk{4} converges by $64^3$, strong-field MHD
turbulence driven at \pk{8} does not converge until $512^3$.
Although very high resolutions are needed to capture the inertial
range of the turbulence, lower resolutions are often adequate for
studying the energy dissipation characteristics of the turbulence as
well as other volume-integrated quantities.

At high Mach number, the ratio of the dissipation timescale to the
flow crossing time at the driving scale increases with increasing
magnetic field strength; however, it does not exceed unity even for
strong-field MHD turbulence.  This ratio is independent of driving
scale.  The fractions of the kinetic energy in solenoidal and
compressive modes are also independent of driving scale, with the
compressive fraction generally being more than twice as high for
hydrodynamic than for strong-field MHD turbulence.


Our spherically-integrated velocity power spectra for decaying,
subsonic hydrodynamic turbulence show evidence of the bottleneck
effect in the velocity power spectrum, consistent with the findings
of Sytine et al.~(2000).  We find more power at high wavenumber in
our driven, supersonic hydro and MHD turbulence simulations than was
found by V03, but it is unclear, particularly in the MHD case, whether
or not this is due to a bottleneck.  Resolutions exceeding $1024^3$
will be necessary to draw firm conclusions about the slope of the
inertial range.  The cylindrically-averaged velocity power spectrum
for driven MHD turbulence is very anisotropic; it has a slope that
approximates that of the hydrodynamic case parallel to the magnetic
field, while perpendicular to the field it is much more shallow.  The
compressive component of velocity has an isotropic power spectrum,
contrary to what was found in V03.

We find the compressive component of velocity in driven, supersonic
MHD turbulence to have a power spectrum that is difficult to
distinguish from the velocity spectrum of an initially-spherical MHD
blast wave.  This calls into question the long-held assumption that
supersonic turbulence power spectra result from an energy cascade
facilitated by interactions local in Fourier space.  The analysis of
structure functions may be useful in determining the source of the
power law spectrum in supersonic turbulence, either a Fourier-space
cascade as in incompressible turbulence, or an ensemble of shocks as
in Burgers turbulence.


For hydrodynamic turbulence, we find a power law scaling of the
velocity dispersion with spatial scale, $\sigma(l) \propto l^{0.58}$,
for scales where the velocity dispersion is supersonic.  However, we
find no such power law for strong-field MHD turbulence, where the
sonic scale is not the only scale of interest.  In this case, we find
that the velocity dispersion drops off more rapidly than a power law
as one approaches smaller scales.


We see time-variability in the saturation energies comparable to that
of the equivalent runs in S98 despite the impulsive driving employed
therein.  Our time-variability increases when we apply our turbulent
driving at larger scales, but remains lower than that shown in K07
even with \pk{2}.  It is possible that the acceleration-based driving
method of K07 is responsible for the difference.  At this large
driving scale, the time-variability increases with Mach number for
both hydro and strong-field MHD.  The method used to drive the
turbulence appears to have a substantial impact on the resulting
turbulent state.

Further investigation should be conducted to determine how much of the
increase in time-variability with driving scale is due to the limited
range of scales over which an inverse cascade can occur.  If the level
of time-variability is determined to be a result of simulation setup
and not representative of the physical system we are trying to
simulate, better diagnostics or simulation methods need to be
developed in order to quantify intermittency in turbulent media.
Although we have very poor statistics in our temperature intermittency
analysis, it appears that this, too, increases with driving scale.
Considering the difference that the driving method makes on the
results, it may be more realistic to study decaying turbulence instead
of driving it arbitrarily.


For identical turbulent data cubes, we find that the post-processing
methods chosen significantly influence the results.  The turbulent
Mach number changes by $\sim 4\%$ depending on the method used in its
computation.  The means of computing the power spectrum also has a
strong influence on the power at wavenumbers overlapping with the
inertial range.  When comparing results published by different groups,
one should keep in mind that this, as well as the code, driving
method, and initial conditions, can affect the quantities being
compared.


Although the saturation energies and energy dissipation characteristics
of the turbulence converge at resolutions within our current
computational capabilities, the power spectra appear to require much
higher resolutions to provide valuable information.  Also considering
that turbulent power spectra can be approximated by non-turbulent
phenomena, it would seem that, for the time being, our focus should be
on other diagnostics.


In conducting these numerical simulations, we have made many
simplifications in order to make the problem more tractable.  These
assumptions, however, may prove to significantly impact the results.
Future studies should consider non-ideal MHD in order to model
low-ionization molecular clouds, a non-isothermal equation of state in
order to study heating and cooling, and self-gravity in order to
follow the collapse of the bound clumps that form in the turbulent
medium.

\acknowledgements

We thank Eve Ostriker for very productive discussion and Tom Gardiner
for his invaluable assistance in bringing Athena up to the task of
successfully simulating supersonic turbulence.  We would also like to
thank Alexei Kritsuk, Cristoph Federrath, and Alex Lazarian for their
feedback.  Simulations were performed on the IBM Blue Gene at
Princeton and on computational facilities supported by NSF grants
AST-0216105 and AST-0722479.

\appendix

\section{Supersonic Turbulence with Godunov Schemes}\label{sec:godunov}

The calculations presented in this paper were conducted using Athena,
a directionally-unsplit, higher-order Godunov code.  This code exactly
conserves mass, momentum, and magnetic flux, as well as energy when assuming
an adiabatic equation of state.  The code captures shocks well and has
a low level of numerical dissipation.  Although the full details of
the algorithms and implementation can be found in the literature
(Gardiner \& Stone 2005, 2008; Stone et al.~2008; Stone \& Gardiner
2008), we will briefly summarize here, noting any modifications made
to the algorithms in order to successfully run the challenging problem
of high Mach number turbulence.

The integration algorithm used in our calculations is a simple
predictor-corrector scheme based on the MUSCL-Hancock scheme of van
Leer (2006), combined with the constrained transport method of Evans
\& Hawley (1998) to enforce the divergence-free constraint on the
magnetic field (i.e. the VL+CT algorithm described in detail in
Stone \& Gardiner 2008).  We find the additional diffusion associated
with this scheme as compared to our CTU+CT algorithm (described in
detail in Stone et al.~2008) to make it more robust to the strong
rarefactions that arise in a highly turbulent medium.  Although the
algorithm is formally only second-order accurate, we use third order
(piecewise parabolic) spatial reconstruction, finding it to provide
more accurate solutions in test problems due to smaller truncation
error.

For our isothermal hydrodynamics runs, we found there to be strong
rarefactions within the turbulent medium for which an approximate
Riemann solver simply was not accurate enough, necessitating the
use of an exact nonlinear Riemann solver.  For adiabatic
hydrodynamics as well as isothermal and adiabatic MHD, we were able to
use approximate nonlinear Riemann solvers, namely HLLC for the hydro
case and HLLD for MHD, more details about which can be found in Stone
et al.~(2008).  Although we used our own adaptation of the adiabatic
HLLD solver of Miyoshi \& Kusano (2005) for our isothermal MHD runs,
we found it to produce turbulent states extremely similar to those
from the isothermal HLLD solver of Mignone (2007).

Although these Riemann solvers are positive definite in 1D, it is not
guaranteed that they will be so in multidimensions.  In fact, we found
that, under extreme conditions, the net mass flux out of a cell in our
isothermal 3D turbulent medium occasionally exceeded the cell's
initial mass.  In the rare circumstance that this occurred, we
recomputed the fluxes of all variables through all interfaces
bordering such cells using first-order reconstruction.  In our Mach 7
strong-field MHD run at $512^3$ that uses \pk{2}, this affected only
a fraction $3 \times 10^{-10}$ of the fluxes computed in the
corrector step of the integrator.  Dropping to first order introduced
enough diffusion in the immediate vicinity to keep the cell-averaged
density positive, while having a negligible effect on the overall
system.  Adding diffusion in this manner instead of enforcing a
density floor maintains exact conservation.

\end{document}